\documentclass[twocolumn,nofootinbib,amsmath,amssymb,a4paper,superscriptaddress]{revtex4}

\usepackage{graphicx}
\usepackage{dcolumn}
\usepackage{bm}

\newcommand{\ket}[1]{\ensuremath{| {#1} \rangle }}
\newcommand{\bra}[1]{\ensuremath{\langle {#1} |}}

\renewcommand{\vec}[1]{\bm{#1}}

\begin{document}

\title{On the extraction of spectral densities from lattice correlators}

\author{Martin Hansen}
\affiliation{INFN Roma Tor Vergata,
Via della Ricerca Scientifica 1, I-00133, Rome, Italy.}

\author{Alessandro Lupo}
\affiliation{University of Rome Tor Vergata,
Via della Ricerca Scientifica 1, I-00133, Rome, Italy.}

\author{Nazario Tantalo}
\affiliation{University of Rome Tor Vergata and INFN Roma Tor Vergata,
Via della Ricerca Scientifica 1, I-00133, Rome, Italy.}

\begin{abstract}
Hadronic spectral densities are important quantities whose non--perturbative knowledge allows for calculating phenomenologically relevant observables, such as inclusive hadronic cross--sections and non--leptonic decay--rates. The extraction of spectral densities from lattice correlators is a notoriously difficult problem because lattice simulations are performed in Euclidean time and lattice data are unavoidably affected by statistical and systematic uncertainties. In this paper  
we present a new method for extracting hadronic spectral densities from lattice correlators. The method allows for choosing a smearing function at the beginning of the procedure and it provides results for the spectral densities smeared with this function together with reliable estimates of the associated uncertainties. The same smearing function can be used in the analysis of correlators obtained on different volumes, such that the infinite volume limit can be studied in a consistent way. While the method is described by using the language of lattice simulations, in reality it is completely general and can profitably be used to cope with inverse problems arising in different fields of research.   
\end{abstract}

\maketitle

\section{introduction}
Hadronic spectral densities are crucial ingredients in the calculation of physical observables associated with the continuum spectrum of the QCD Hamiltonian. A notable classical example is provided by the differential cross section for the process $e^+ e^- \mapsto \text{hadrons}$ that, at leading order in the electromagnetic coupling, is proportional to the QCD spectral density evaluated between hadronic electromagnetic currents,
\begin{flalign} 
\frac{d\Sigma(E)}{dE}\propto \bra{0}J_{em}^k(0)\, \delta(H-E)\delta^3(\vec P)\, J_{em}^k(0)\ket{0}\;,
\label{eq:starting}
\end{flalign}
where $E$ is the energy of the electron--positron pair in the center of mass frame, $H$ and $\vec P$ are respectively the QCD Hamiltonian and total momentum operators and $J_{em}^\mu(x)$ is the hadronic electromagnetic current. Other important examples of observables, in which spectral densities play a crucial r\^ole, are the flavour--changing non--leptonic decay--rates of kaons and heavy flavoured mesons, the deep inelastic scattering cross--section, and thermodynamic observables arising in the study of QCD at finite--temperature and of the quark--gluon plasma.

It is notoriously difficult to obtain model--independent non--perturbative theoretical predictions for hadronic spectral densities. In principle this is a problem that can be addressed from first--principles within the solid framework of lattice QCD. However, in practice, one has to face highly non--trivial numerical and theoretical problems in order to extract spectral densities from lattice simulations. 

The origin of these problems can be traced back to the fact that lattice results unavoidably are affected by statistical and systematic errors. More precisely, the primary observables computed in a lattice simulation are Euclidean time--ordered correlators at discrete values of the space--time coordinates and on a finite volume, e.g.
\begin{flalign}
C(t)=\frac{1}{L^3}\sum_{\vec x}\, T\bra{0} O(x)\, \bar{O}(0) \ket{0}_L\;,
\end{flalign}
where $L$ is the linear extent of the spatial volume $V=L^3$ while $O$ and $\bar O$ are generic hadronic operators. In the following we shall not discuss cutoff effects and, therefore, we shall not indicate the dependence of the different quantities upon the lattice spacing. We shall however always assume that the correlators are known only for discrete values of the space--time coordinates. At positive Euclidean times $t\ge 0$ the previous correlator can be rewritten as 
\begin{flalign}
C(t)=\int_0^\infty\, dE\, \rho_L(E)\, e^{-tE}\;,
\end{flalign}
where, for simplicity (see below for a generalization in the case of periodic boundary conditions in time), we have assumed that the time extent of the lattice is infinite and where we have defined the associated spectral density
\begin{flalign}
\rho_L(E)=\frac{1}{L^3}\sum_{\vec x}\, \bra{0} O(0,\vec x)\, \delta(E-H_L)\, \bar{O}(0) \ket{0}_L\;.
\end{flalign}
The main problems faced during the extraction of spectral densities from lattice simulations can now be explained by starting from the previous two expressions. 

The first problem is associated with the fact that the extraction of $\rho_L(E)$ from the measured lattice correlator $C(t)$ requires an inverse Laplace--transform to be performed numerically, an ill--posed problem when the measured data are affected by uncertainties. This is the case for $C(t)$ that unavoidably will be affected by statistical errors, particularly at large time separations where (a part from some countable exceptions) the signal--to--noise ratio degrades exponentially for Euclidean hadronic correlators. 

The second problem comes from the fact that the finite volume Hamiltonian $H_L$ has a discrete spectrum. The finite volume spectral density is a distribution having support in correspondence with the eigenvalues $E_n(L)$ of $H_L$,
\begin{flalign}
\rho_L(E)=\sum_{n}\, w_n(L)\, \delta(E-E_n(L))\;.
\end{flalign}
Here it is important to notice that the finite volume spectral densities cannot be directly associated, even in the ideal case in which these can be computed exactly, to physical observables.  

A big step forward in the extraction of spectral densities from finite--volume lattice simulations has recently been done in Ref.~\cite{Hansen:2017mnd} where it has been suggested to rely on a method originally devised to analyse geophysical observations in Ref.~\cite{Backus}. The central idea of the so--called Backus--Gilbert method\footnote{See the references quoted in~\cite{Hansen:2017mnd} for previous applications of the Backus--Gilbert method in the context of lattice simulations and Refs.~\cite{Asakawa:2000tr,Burnier:2013nla,Rothkopf:2016luz} for an incomplete list of references on other approaches used to extract spectral densities from lattice correlators.} (see also Ref.~\cite{numrec}) is to optimize the amount of information that can be extracted from noisy measurements, in our case $C(t)$, by focusing on the calculation of \emph{smeared} spectral densities,
\begin{flalign}  
\hat \rho_L(\sigma,E_\star)=\int_0^\infty dE\, \Delta_\sigma(E_\star,E)\, \rho_L(E)\;. 
\end{flalign}  
Notice that smeared spectral densities are smooth functions of the energy (as opposed to the distributions $\rho_L(E)$) and that the study of their infinite--volume limit at fixed smearing function is a well--posed problem. Ideally one would like to choose the smearing functions $\Delta_\sigma(E_\star,E)$ with support in a region around $E_\star$ of width proportional to $\sigma$ and such that they become Dirac $\delta$--functions in the limit in which the smearing radius parameter $\sigma$ is sent to zero. If one can choose the same smearing function on different volumes, the infinite--volume spectral density can then be extracted by taking the double limit  
\begin{flalign}  
\rho(E_\star) = \lim_{\sigma\to 0}\, \lim_{L\to \infty}\, \hat \rho_L(\sigma,E_\star)~,
\end{flalign}  
in the specified order. 

As already done by the authors of Ref.~\cite{Hansen:2017mnd}, we stress that the limit of vanishing smearing radius might not be necessary in order to compare theoretical predictions with experimental data. Indeed, it might be possible to smear experimental observations with the same function used in the theoretical calculation. In the case of $e^+ e^- \mapsto \text{hadrons}$ one should for example compare the theoretical predictions with $\int_0^\infty d\Sigma(E)\, \Delta_\sigma(E_\star,E)$. In fact one should also notice that, in order to derive results such as Eq.~(\ref{eq:starting}), a smearing function \emph{has} to be introduced at intermediate steps of the calculations, after which the limit of vanishing smearing radius has to be taken. This point is extensively discussed in Ref.~\cite{Hansen:2017mnd} by making contact with the so--called Fermi's golden rule.

On the one hand, the Backus--Gilbert method is an extremely efficient algorithm for controlling the statistical errors on the estimated smeared spectral densities. On the other hand, the shape of the smearing function cannot be chosen arbitrarily in this method, it is an \emph{output} of the procedure. Indeed, the algorithm is designed in such a way that the width of the smearing function (having the properties of being peaked around $E_\star$ and of having unit area) is optimized on the basis of the number of observations and of their statistical uncertainties (in our case the number of discrete times $t$ at which $C(t)$ and the associated statistical errors are known). This feature of the algorithm may not represent a problem in experimental applications of the Backus--Gilbert method because, at the end of the procedure, the resulting smearing function is known and no infinite--volume limit has to be taken. It is instead a strong limitation in the context of lattice simulations where simulations at different volumes produce results with different statistical uncertainties and different number of points. In this case one gets different smearing functions at different volumes and the extraction of the infinite--volume physical observable becomes, if not impossible, extremely difficult.  

In this paper we present a new method, a generalization of the Backus--Gilbert approach, in which the shape of the smearing function is an \emph{input} of the procedure and not an output. The method uses the same mechanism of the original Backus--Gilbert proposal to keep the statistical errors under control. This happens at the price of a distortion of the target smearing function induced by the presence of statistical errors and by the finite number of observations. At the end of the numerical procedure the systematic error associated with this distortion can be reliably quantified and added to the statistical uncertainties in order to provide a reliable estimate of the smeared spectral densities. Our method gives an exact reconstruction of the spectral densities, smeared with the chosen functions, in the limit of vanishing statistical uncertainties and of an infinite number of discrete lattice points along the time direction. The method is quite general and we are pretty confident that it will be useful in addressing ``inverse problems'' arising in different fields of research, for example the problems where the application of the Backus--Gilbert method has already proven to be useful.    
 
The rest of the paper is organized as follows. In section~\ref{sec:BGreview} we review of the original Backus--Gilbert method and in section~\ref{sec:ourmethod} we present our method. In section~\ref{sec:benchmark} we apply our method to a benchmark system where we know the exact spectral density, while in section~\ref{sec:lqcd} we apply the method to real lattice correlators. We draw our conclusions in section~\ref{sec:conclusions}. The paper ends with two appendices. Appendix~\ref{sec:appendix} contains the explicit formulae needed to implement our method in computer programs. Appendix~\ref{sec:appendix2} contains additional examples of applications of our method in the case of synthetic data.

\section{Review of the Backus--Gilbert method}
\label{sec:BGreview}
In this section we review the original Backus--Gilbert method. This will help us to set the notation and to discuss, in the following sections, the similarities and the differences between our new proposal and the Backus--Gilbert approach. Although the method is general and can be applied to many different problems, in our discussion we shall use the language of lattice correlators to explain the approach. The generalization to other contexts is straightforward, see Ref.~\cite{numrec}.

Let us consider a generic lattice correlator that, by following the same steps of the introduction, can be written as
\begin{flalign}
C(t)=\int_0^\infty\, dE\, \rho_L(E)\, b_T(t,E)\;,
\end{flalign}
where $\rho_L(E)$ is the distribution corresponding to the finite volume spectral density. Here $b_T(t,E)$ are the so--called basis functions. These are simply given by exponentials in the limit of an infinite temporal extent of the lattice,
\begin{flalign} 
b_{\infty}(t,E)= e^{-tE}\;, 
\label{eq:basis1}
\end{flalign} 
while, in the case of periodic boundary conditions in time and by assuming that the correlator is symmetric under time--reversal, the basis functions are\footnote{The use of Eq.~(\ref{eq:basis2}) in the case of periodic boundary conditions in time is an approximation. The approximation is much better than the naive use of Eq.~(\ref{eq:basis1}) at finite values of $T$ but the general spectral decomposition of a periodic hadronic correlator would require the inclusion of other contributions that vanish exponentially fast when $T$ is sent to infinity. See Ref.~\cite{DelDebbio:2007pz} for a discussion of this point.}
\begin{flalign} 
b_{T}(t,E)= e^{-tE}+e^{-(T-t)E}\;.
\label{eq:basis2}
\end{flalign} 
In all what follows the time variable $t$ is assumed to be discrete, non negative and smaller than the time extent of the lattice ($0\le t< T$). 

The central idea of the Backus--Gilbert method is to search for a smearing function that lives in the space spanned by the basis functions, more precisely
\begin{flalign}
\Delta^{BG}(E_\star,E)=\sum_{t=0}^{t_{max}}g_t(E_\star)\, b_{T}(t+1,E)\;,
\label{eq:bgdeltadef}
\end{flalign}
with $t_{max}<T/2$. Once the coefficients $g_t(E_\star)$ that define the smearing function are known, the smeared spectral density can then easily be computed by starting from the correlator,
\begin{equation}
\begin{split}
\hat \rho_L^{BG}(E_\star)
&=\sum_{t=0}^{t_{max}}g_t(E_\star)\, C(t+1) \\
&= \int_0^\infty\, dE\, \rho_L(E)\, \Delta^{BG}(E_\star,E)\;.
\label{eq:bgreconstruction}
\end{split}
\end{equation}

The Backus--Gilbert procedure ``optimizes'' the choice of the smearing function, i.e. of the coefficients $g_t(E_\star)$, on the basis of the measured data for the correlator. In the absence of statistical errors the coefficients are fixed by minimizing a deterministic functional that can be interpreted as a measure of the width of the smearing function. The functional is
\begin{flalign}
A_{BG}[g]= \int_0^\infty\, dE\, (E-E_\star)^2\, \left\{\Delta^{BG}(E_\star,E)\right\}^2~,
\end{flalign}
and the minimization is performed under the unit area constraint
\begin{flalign}
\int_0^\infty\, dE\, \Delta^{BG}(E_\star,E)=1\;.
\label{eq:unitareabg}
\end{flalign}
It is a simple exercise (see Ref.~\cite{numrec}) to show that the solution of this problem is given by
\begin{flalign}
\vec g(E_\star) = \frac{\mathtt{A}^{-1}(E_\star)\, \vec R }{\vec R^T\, \mathtt{A}^{-1}(E_\star)\, \vec R }\;,
\label{eq:bggnoerror}
\end{flalign}
where we have used a vector notation for the coefficients, $\vec g^T(E_\star)=(g_0(E_\star),\cdots,g_{t_{max}}(E_\star))$, the vector $\vec R$ has the following entries
\begin{flalign}
R_t = \int_0^\infty\, dE\, b_T(t+1,E)
\label{eq:rdef}
\end{flalign}
and the elements of the matrix $\mathtt{A}(E_\star)$ are given by
\begin{flalign}
\mathtt{A}_{tr}(E_\star)= \int_0^\infty\, dE\, (E-E_\star)^2\, b_T(t+1,E)\, b_T(r+1,E)\;.
\label{eq:abg}
\end{flalign}
It is important to notice that the matrices $\mathtt{A}(E_\star)$ tend to be nearly singular for the basis functions discussed in this paper. From a numerical point of view this might be an issue, but in fact the problem can easily be circumvented on currently available computers by using extended--precision arithmetic. In order to avoid coping with algorithmic instabilities induced by numerical rounding errors, this is what we have done in our computer programs~\cite{code}. 

\begin{figure}[!t]
\includegraphics[width=0.48\textwidth]{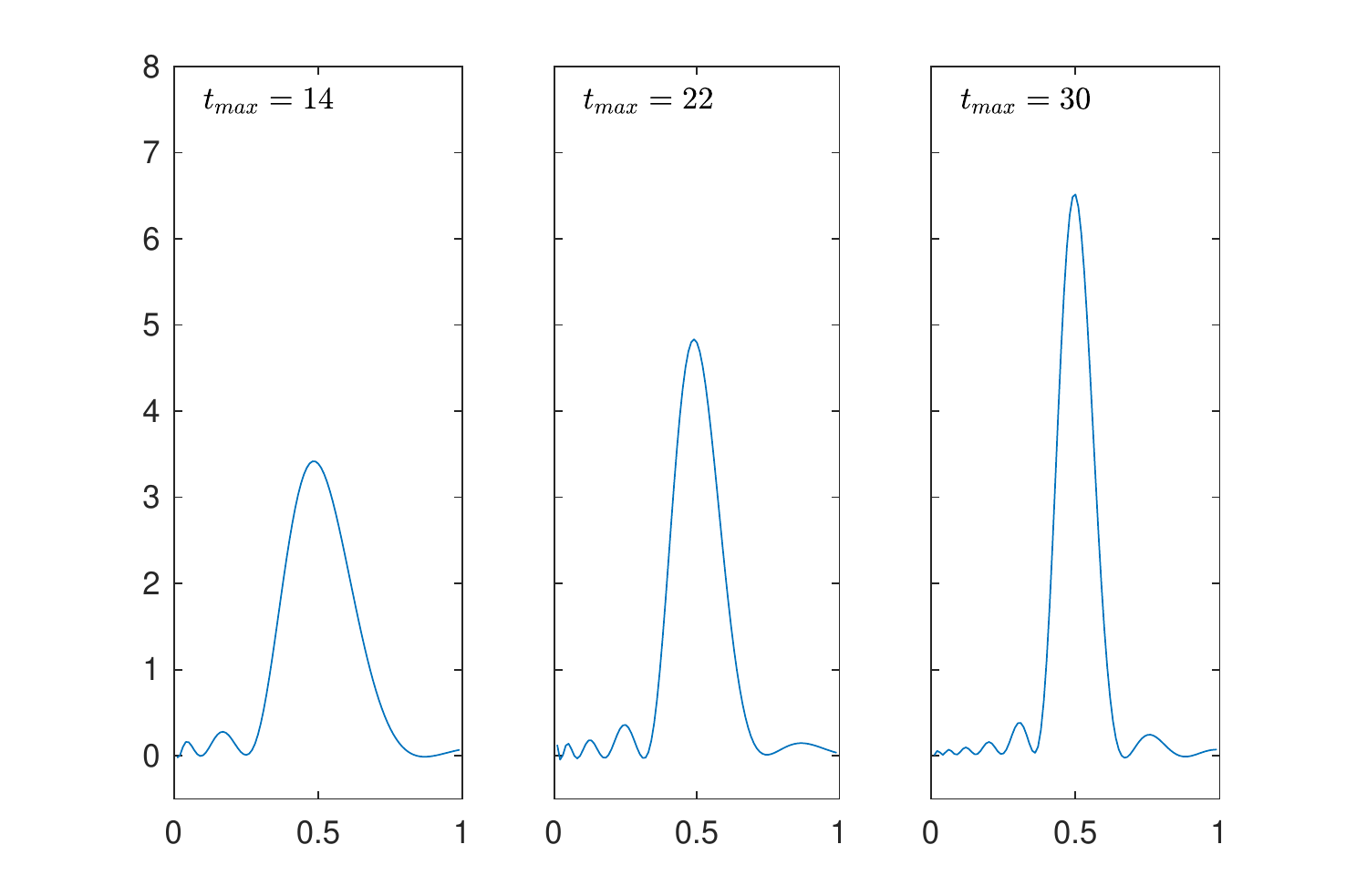}
\caption{\label{fig:bgfunctions} 
Smearing functions $\Delta^{BG}(E_\star,E)$ obtained by applying the Backus--Gilbert procedure in the absence of statistical errors with $E_\star=0.5$ and
$b_{\infty}(t,E)$ as basis functions. The different panels correspond to different values of $t_{max}$. As it can be seen the function $\Delta^{BG}(E_\star,E)$ gets more similar to a Dirac $\delta$--function for increasing values of $t_{max}$.}
\includegraphics[width=0.48\textwidth]{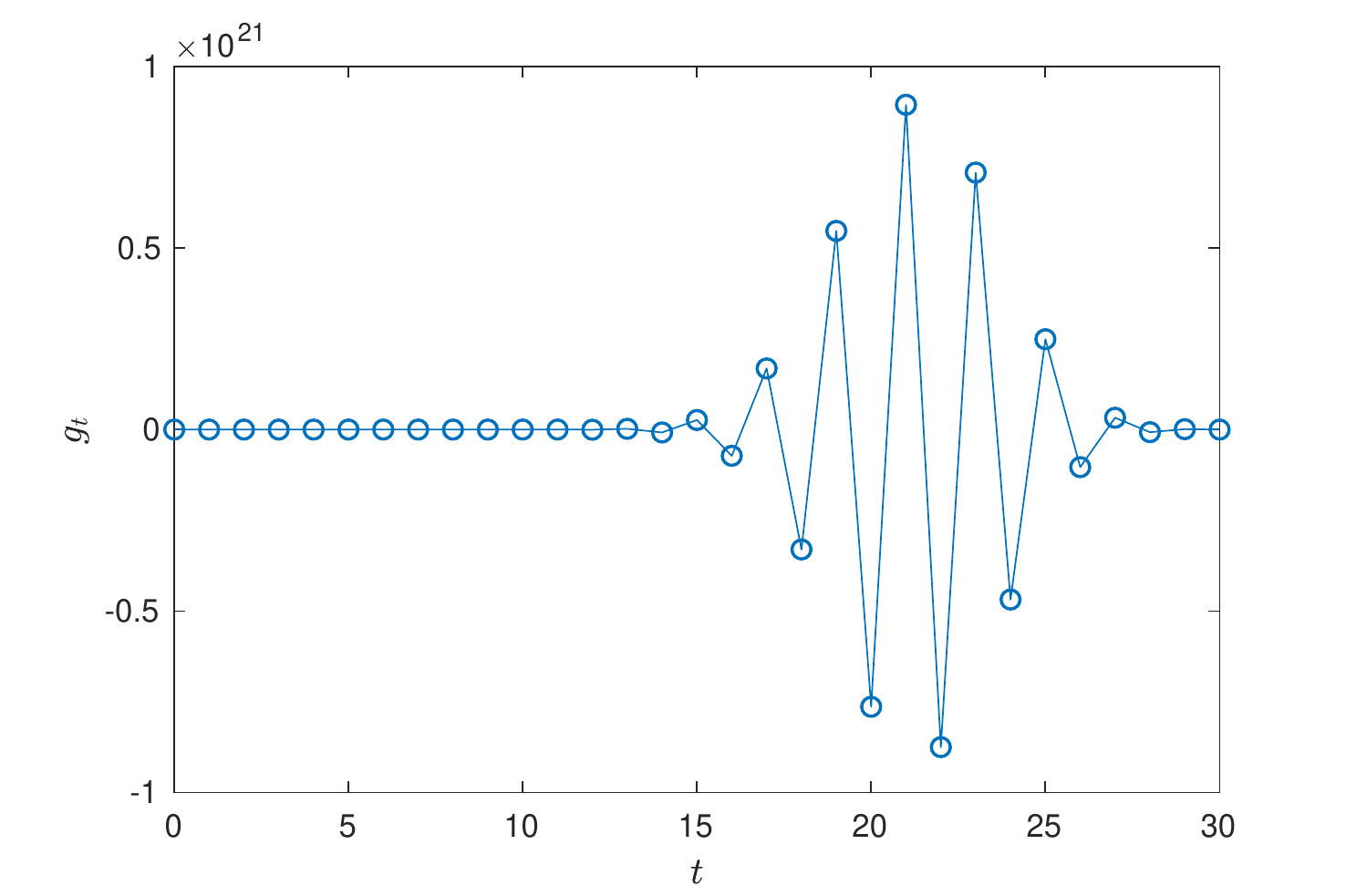}
\caption{\label{fig:bgcoefficients} 
Values of the coefficients $g_t(E_\star)$ corresponding to the smearing function $\Delta^{BG}(E_\star,E)$ shown in the right--panel of Figure~\ref{fig:bgfunctions}, i.e. the coefficients obtained by applying the Backus--Gilbert procedure in the absence of statistical errors with $E_\star=0.5$, $b_{\infty}(t,E)$ as basis functions and $t_{max}=30$. A typical pattern for these coefficients is that they change sign and for some values of $t$ they have extremely large absolute values (the scale on the $y$--axis varies between $\pm 10^{21}$).}
\end{figure}
In Figure~\ref{fig:bgfunctions} we show some examples of the smearing functions obtained by using Eq.~(\ref{eq:bggnoerror}). As it can be seen by comparing the plots in the different panels, the function $\Delta^{BG}(E_\star,E)$ becomes more similar to a Dirac $\delta$--function when increasing the number of basis functions used in its definition. 

In Figure~\ref{fig:bgcoefficients} we show the coefficients $g_t(E_\star)$ corresponding to the smearing function $\Delta^{BG}(E_\star,E)$ shown in the right--panel of Figure~\ref{fig:bgfunctions}. The plot has been shown in order to highlight a typical feature exhibited by these coefficients. As a consequence of the nearly singular nature of the matrix $\mathtt{A}(E_\star)$ the coefficients become gigantic for some values of $t$ and, moreover, oscillate in sign. Having noticed this feature we can now discuss the Backus--Gilbert procedure in the presence of uncertainties. 

The exact correlator is an idealization that is not accessible in the real world and, in the presence of (experimental or) statistical errors, we have to consider
\begin{flalign}
C_i(t)=\bar C(t) + \delta C_i(t)\;,
\qquad
i=0,\dots, N-1
\end{flalign}
where the index $i$ runs over the $N$ different statistical samples (for a lattice correlator we can think to the different bootstrap or jackknife bins), $\bar C(t)$ is the statistical average and $\delta C_i(t)$ is the deviation from the average of the $i$-th bin, $\sum_{i=0}^{N-1}\delta C_i(t)=0$. 

Given the fact that the coefficients $g_t(E_\star)$ are huge numbers, even a tiny deviation (for example an apparently harmless rounding error) from the average gives an unacceptably large contribution to the smeared spectral function. Indeed, by applying Eq.~(\ref{eq:bgreconstruction}) to $C_i(t)$, one gets that the sums $\sum_t g_t(E_\star)\, \delta C_i(t)$ are also huge numbers in general and the final error on the estimated smeared spectral functions turns out to be unacceptably large. This can be viewed as a manifestation of the fact that we are dealing with a numerically ill--posed problem.

In order to keep statistical errors under control Backus and Gilbert considered another functional of the coefficients, a measure of the statistical error on the smeared spectral function, namely
\begin{flalign}
B[g]= \vec g^T\, \mathtt{Cov}\, \vec g\;,
\end{flalign}
where $\mathtt{Cov}$ is the covariance matrix of the correlator, 
\begin{flalign}
\mathtt{Cov}_{tr} = \frac{1}{N-1}\sum_{i=0}^{N-1} \delta C_i(t+1)\delta C_i(r+1)\;.
\end{flalign}
In the presence of statistical errors, the coefficients are fixed by minimizing the following functional
\begin{flalign}
W[\lambda,g]=(1-\lambda)A_{BG}[g] +\lambda B[g]\;,
\end{flalign}
again under the unit area constraint of Eq.~(\ref{eq:unitareabg}). In this case the solution is given by
\begin{flalign}
\vec g(\lambda,E_\star) = \frac{\mathtt{W}^{-1}(\lambda,E_\star)\, \vec R }{\vec R^T\, \mathtt{W}^{-1}(\lambda,E_\star)\, \vec R }\;,
\end{flalign}
where the matrix $\mathtt{W}(\lambda,E_\star)$ has elements
\begin{flalign}
\mathtt{W}_{tr}(\lambda,E_\star)= (1-\lambda) \mathtt{A}_{tr}(E_\star) + \lambda\, \mathtt{Cov}_{tr}
\label{eq:wbg}
\end{flalign}
with $\mathtt{A}_{tr}(E_\star)$ already defined in Eq.~(\ref{eq:abg}). The real number $\lambda$ is a free parameter of the algorithm, chosen in the range $[0,1]$. 

The functional $W[\lambda,g]$ is in fact a convex linear combination of
the deterministic functional $A_{BG}[g]$ and of the error functional $B[g]$. The presence of the error functional in the minimization procedure forbids solutions
corresponding to gigantic values of the coefficients. Statistical errors are thus kept under control at the price of accepting that the shape of the smearing function is determined (somehow optimized) also by the statistical errors. 

The tuning of the parameter $\lambda$ is a subtle issue in the Backus--Gilbert procedure. Choosing $\lambda$ too small\footnote{Notice that our parameter $\lambda$ corresponds to $1-\lambda$ in Ref.~\cite{Hansen:2017mnd}.} may result in too large statistical errors while values of $\lambda$ close to one may generate smearing functions that are useless for physical applications. This point will be discussed in more details in section~\ref{sec:benchmark} where, by using a synthetic correlator generated by starting from an exactly known spectral function and by adding random statistical noise, we shall compare the results obtained with the Backus--Gilbert method with the ones obtained by using our method.

\section{The new method}
\label{sec:ourmethod}
In our method the target smearing function is an \emph{input} of the algorithm. For example, the target smearing function can be chosen as a Gaussian of width $\sigma$, centred at $E_\star$ and normalized to have unit area in the interval $[0,\infty)$,
\begin{flalign}
\Delta_\sigma(E_\star,E) = \frac{e^{-\frac{(E-E_\star)^2}{2\sigma^2}}}{\int_{0}^\infty dE\, e^{-\frac{(E-E_\star)^2}{2\sigma^2}}}\;,
\label{eq:gaussian}
\end{flalign}
The method then searches for an optimal approximation of the target smearing function in the space spanned by the basis functions,
\begin{flalign}
\bar \Delta_\sigma(E_\star,E)=\sum_{t=0}^{t_{max}}g_t(\lambda,E_\star)\, b_{T}(t+1,E)\;,
\label{eq:deltabardef}
\end{flalign}
where $t_{max}<T/2$. The previous formula is identical to the definition of the smearing function in the original Backus--Gilbert proposal, see Eq.~(\ref{eq:bgdeltadef}) above. The difference is in the way the coefficients $g_t(\lambda,E_\star)$ are determined.

This is done by minimizing again a convex linear combination of a deterministic functional and of the error functional,
\begin{flalign}
W[\lambda,g]=(1-\lambda)A[g] +\lambda \frac{B[g]}{C(0)^2}\;,
\end{flalign}
under the unit area constraint
\begin{flalign}
\int_{0}^\infty dE\, \bar \Delta_\sigma(E_\star,E)=1~.
\end{flalign}
However, in our case, the deterministic functional is chosen to be a measure of the difference between the target and the approximated smearing functions, namely
\begin{flalign}
A[g]=
\int_{E_0}^\infty dE\, \left\vert \bar \Delta_\sigma(E_\star,E)-\Delta_\sigma(E_\star,E)\right\vert^2\;,
\end{flalign}
while the error functional is conveniently normalized with $C(0)^2$, i.e. the square of the correlator at $t=0$.

The parameter $E_0$ has to be chosen in such a way that the finite volume spectral function $\rho_L(E_0)=0$. This is always possible in the case of connected correlators in QCD and in the charged sectors of QCD$+$QED because of the presence of a mass gap. According to our experience, having $E_0>0$ is particularly convenient in the case of $b_{T}(t,E)$ as basis functions. 

It is easy to show that the solution of the minimization procedure is given by
\begin{align}
\begin{split}
\vec g(\lambda,E_\star) &= 
\mathtt{W}^{-1}(\lambda) \vec f(\lambda,E_\star) \\
&+\mathtt{W}^{-1}(\lambda)\, \vec R \,
\frac{1-\vec R^T\,\mathtt{W}^{-1}(\lambda)\, \vec f(\lambda,E_\star)}
{\vec R^T\, \mathtt{W}^{-1}(\lambda)\, \vec R }\;,
\label{eq:oursolution}
\end{split}
\end{align}
where the vector $\vec R$ has already been defined in Eq.~(\ref{eq:rdef}) and the components of the vector $\vec f(E_\star)$ are given by
\begin{equation}
f_t(\lambda,E_\star) = (1-\lambda)\int_{E_0}^\infty dE\, b_{T}(t+1,E)\, \Delta_\sigma(E_\star,E)~.
\end{equation}
The matrix $\mathtt{W}$ has the elements
\begin{equation}
 \mathtt{W}_{tr}(\lambda) = (1-\lambda)\mathtt{A}_{tr} + \lambda\frac{\mathtt{Cov}_{tr}}{C(0)^2}~,
\end{equation}
where
\begin{equation}
\mathtt{A}_{tr} = \int_{E_0}^\infty dE\, b_{T}(t+1,E)\,b_{T}(r+1,E)~.
\end{equation}
Explicit expressions for $\vec R$, $\mathtt{A}$ and $\vec f$, derived in the case of the smearing function of Eq.~(\ref{eq:gaussian}), are given in appendix~\ref{sec:appendix}.  

In the absence of statistical errors our procedure is a method to obtain the best approximation of the target smearing function in the space spanned by the basis functions under the norm defined by the functional $A[g]$. Since the target function is assumed to be analytic in the interval $[E_0,\infty)$ and to decay faster than any power for $E$ that goes to infinity, the error of the approximation can be made arbitrarily small by enlarging the space spanned by the basis functions. This can be understood by looking at the deterministic functional in the case of the choice of $b_{\infty}(t,E)$ as basis functions after performing the change of integration variable to $x=e^{-E}$,
\begin{flalign}
A[g]=
\int_0^{e^{-E_0}} dx\, x\left\vert \sum_{t=0}^{t_{max}}g_t x^t-\frac{\Delta_\sigma(E_\star,-\log(x))}{x}\right\vert^2\;.
\end{flalign}
In fact in our procedure we are just searching for the best polynomial approximation of a well--behaved function. The argument holds also in the case where $b_{T}(t,E)$ are the basis functions because these simply reduce to $b_{\infty}(t,E)$ in the limit of infinitely many times\footnote{Notice that $t_{max}<T/2$, see Eq.~(\ref{eq:deltabardef}), so that in order to send $t_{max}$ to infinity one has also to send $T$ to infinity.}.
\begin{figure}[!t]
\includegraphics[width=0.48\textwidth]{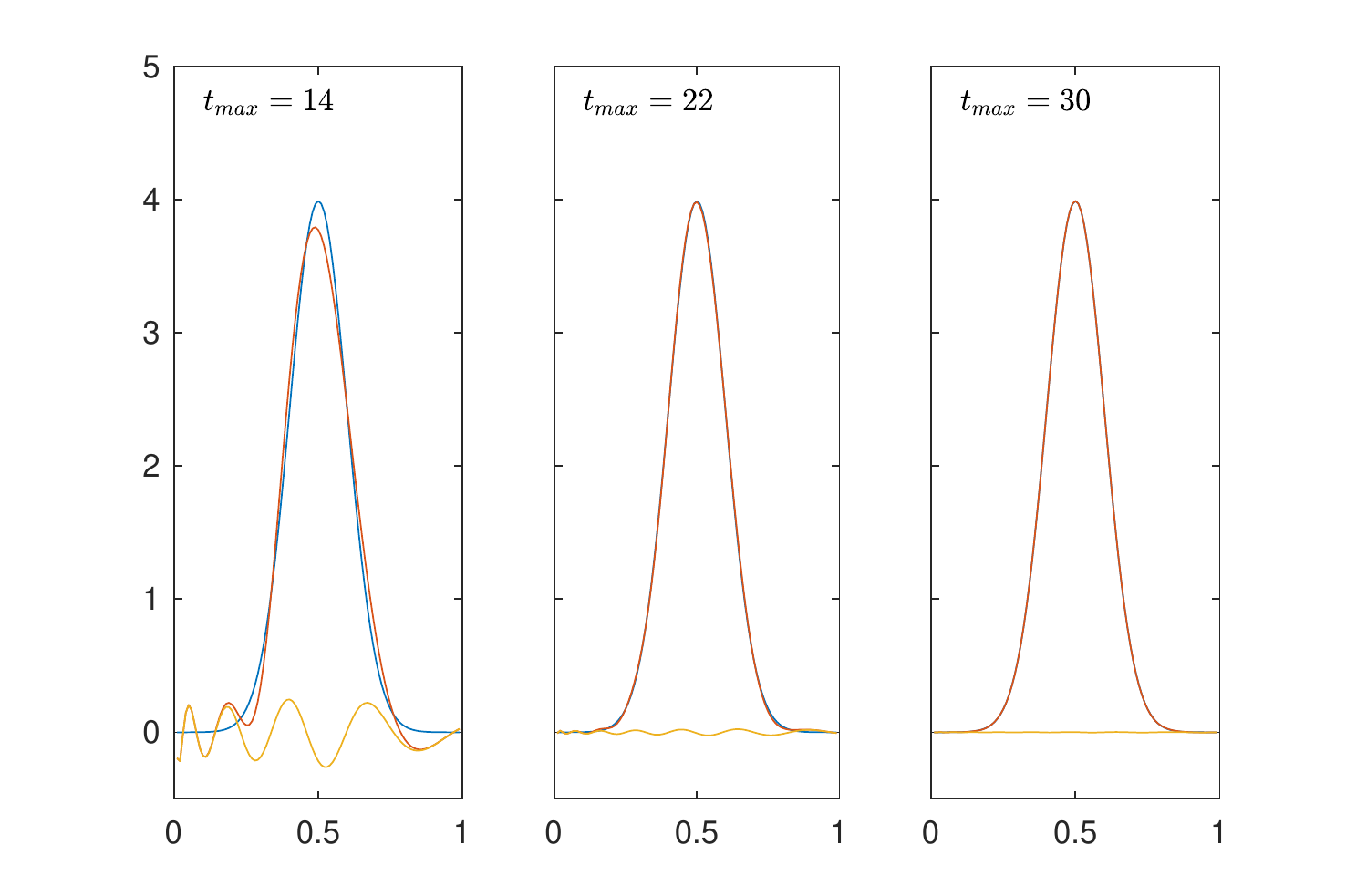}
\includegraphics[width=0.48\textwidth]{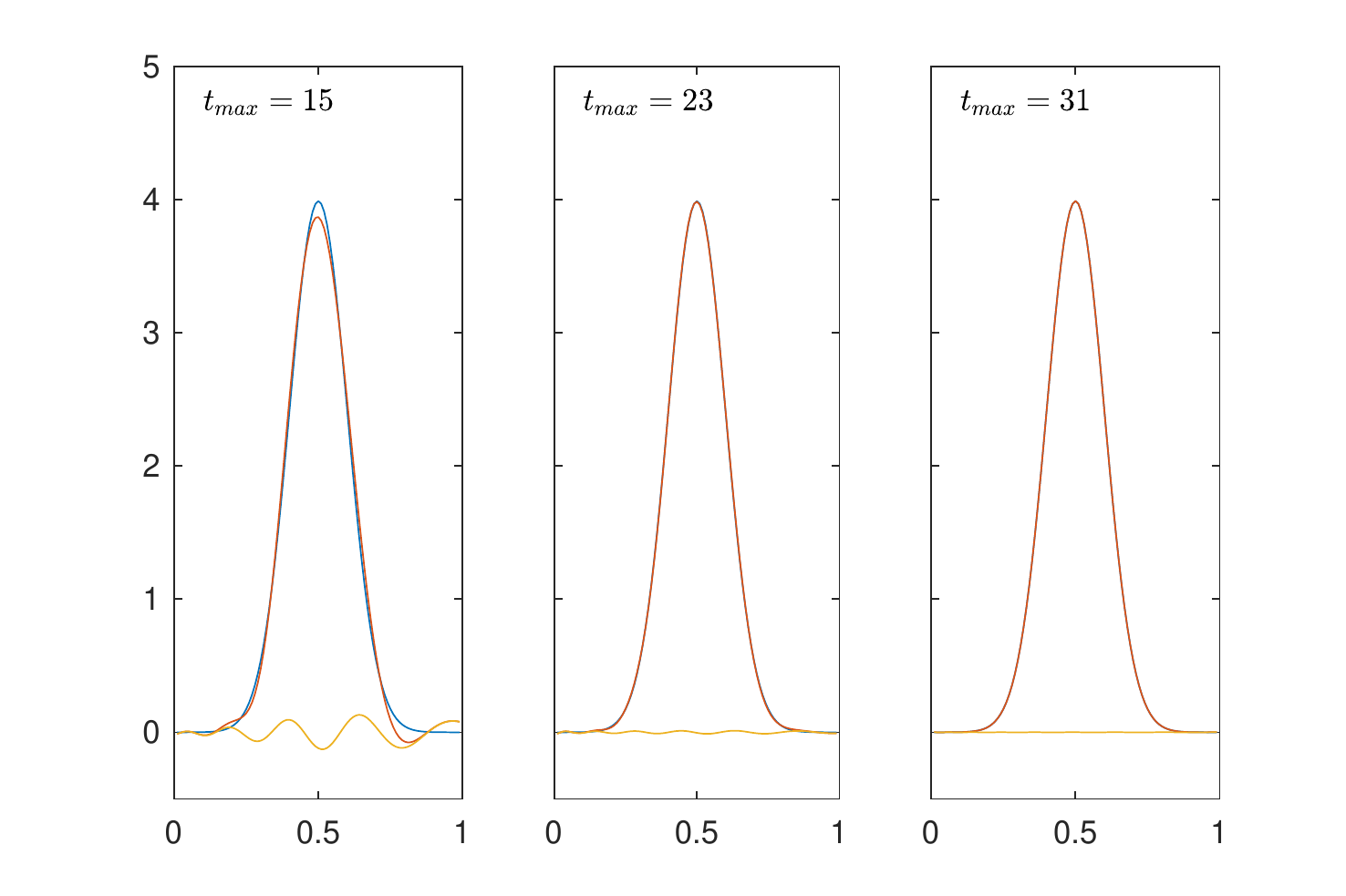}
\caption{\label{fig:ourfunctions}
Comparison of the target smearing function $\Delta_\sigma(E_\star,E)$ (blue curves) at $E_\star=0.5$ and $\sigma=0.1$ with the functions $\bar \Delta_\sigma(E_\star,E)$ (red curves) obtained with our method.
In each row the different panels correspond to different values of $t_{max}$. The panels in the first row correspond to the choice of $b_{\infty}(t,E)$ as basis functions while the panels in the second row to the choice of $b_{T}(t,E)$ with $T=2(t_{max}+1)$. In all plots the yellow curve shows the  difference, and as expected, it goes to zero in the limit of large $t_{max}$.}
\end{figure}
\begin{figure}[!t]
\includegraphics[width=0.48\textwidth]{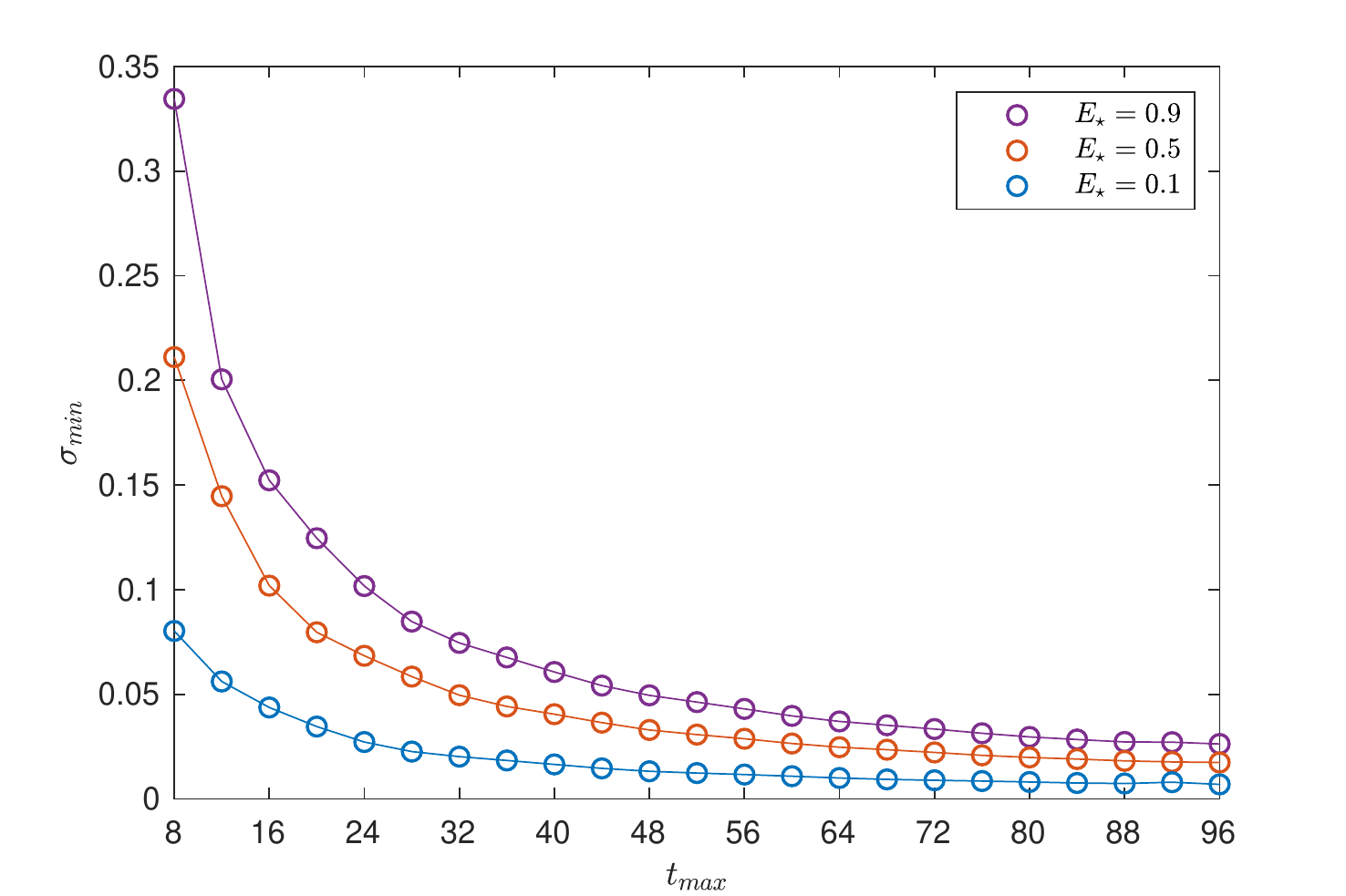}
\caption{\label{fig:sigma_min}Solution to the equation $\delta_\sigma(E_\star,E_\star)=0.05$ as a function of $t_{max}$ for three different values of $E_\star$ in the case where $B[g]=0$. The solution indicates the smallest possible choice of $\sigma$ ensuring that the relative error on the target smearing function is below 5\%.
}
\end{figure}
The comparison of the smearing functions $\bar \Delta_\sigma(E_\star,E)$ obtained with our method in the absence of statistical errors with the target function $\Delta_\sigma(E_\star,E)$ is shown in Figure~\ref{fig:ourfunctions}.

The different philosophy of our method with respect to the original Backus--Gilbert proposal can already be appreciated by comparing Figure~\ref{fig:ourfunctions} and Figure~\ref{fig:bgfunctions}. In the Backus--Gilbert method, by changing $t_{max}$ one gets a different (sharper) function. In our method, by increasing $t_{max}$ one gets a better approximation of the target smearing function. Moreover, in our method the error of the approximation of the target smearing function is known and this information can be used to estimate the final error on the smeared spectral density as we are now going to explain.

On the one hand, in the presence of statistical uncertainties the difference between the target and the approximated smearing functions is due to both a finite value of $t_{max}$ and to the presence of the error functional $B[g]$ in the minimization procedure, whose importance is regulated by the choice of the $\lambda$ parameter. On the other hand, the quantity 
\begin{flalign}
\delta_\sigma(E_\star,E)=1-\frac{\bar \Delta_\sigma(E_\star,E)}{\Delta_\sigma(E_\star,E)}~,
\label{eq:smalldelta}
\end{flalign}
can always be calculated at the end of the procedure.  In Figure~\ref{fig:sigma_min} we show how this quantity depends on the different parameters by solving the equation $\delta_\sigma(E_\star,E_\star)=0.05$ for different choices of $E_\star$ and $t_{max}$. The solution indicates the smallest possible choice of $\sigma$ ensuring that the relative error on the target smearing function is less than 5\% at the peak. In the original Backus-Gilbert method, the choice of $\sigma$ is automatically optimized, but in our method a scan like this can be used for choosing an optimal value for the smearing parameter.

In principle, by knowing the quantity $\delta_\sigma(E_\star,E)$ one can write an exact expression for the bias on the smeared spectral density associated with our method,
\begin{flalign}
&\Delta^{bias} =
\int_{0}^{\infty}dE\, \delta_\sigma(E_\star,E)\, \Delta_\sigma(E_\star,E)\, \rho_L(E)
\;.
\end{flalign}
In practice we cannot use the previous formula for the obvious reasons that we do not know the true spectral density and that we cannot explore the full energy range $[0,\infty)$. The quality of the results obtained with our method can nevertheless be assessed by monitoring the relative deviation $\delta_\sigma(E_\star,E)$. This will be illustrated in the next section where, by using a benchmark system where the exact spectral density is known, we will show that a trustable estimate of the systematic error associated with our method can be obtained by using the formula
\begin{flalign}
\Delta^{syst} =
\left\vert \delta_\sigma(E_\star,E_\star)\right\vert\, \hat \rho_L(\sigma,E_\star)\;.
\label{eq:systerror}
\end{flalign}
\begin{figure}[!t]
\includegraphics[width=0.48\textwidth]{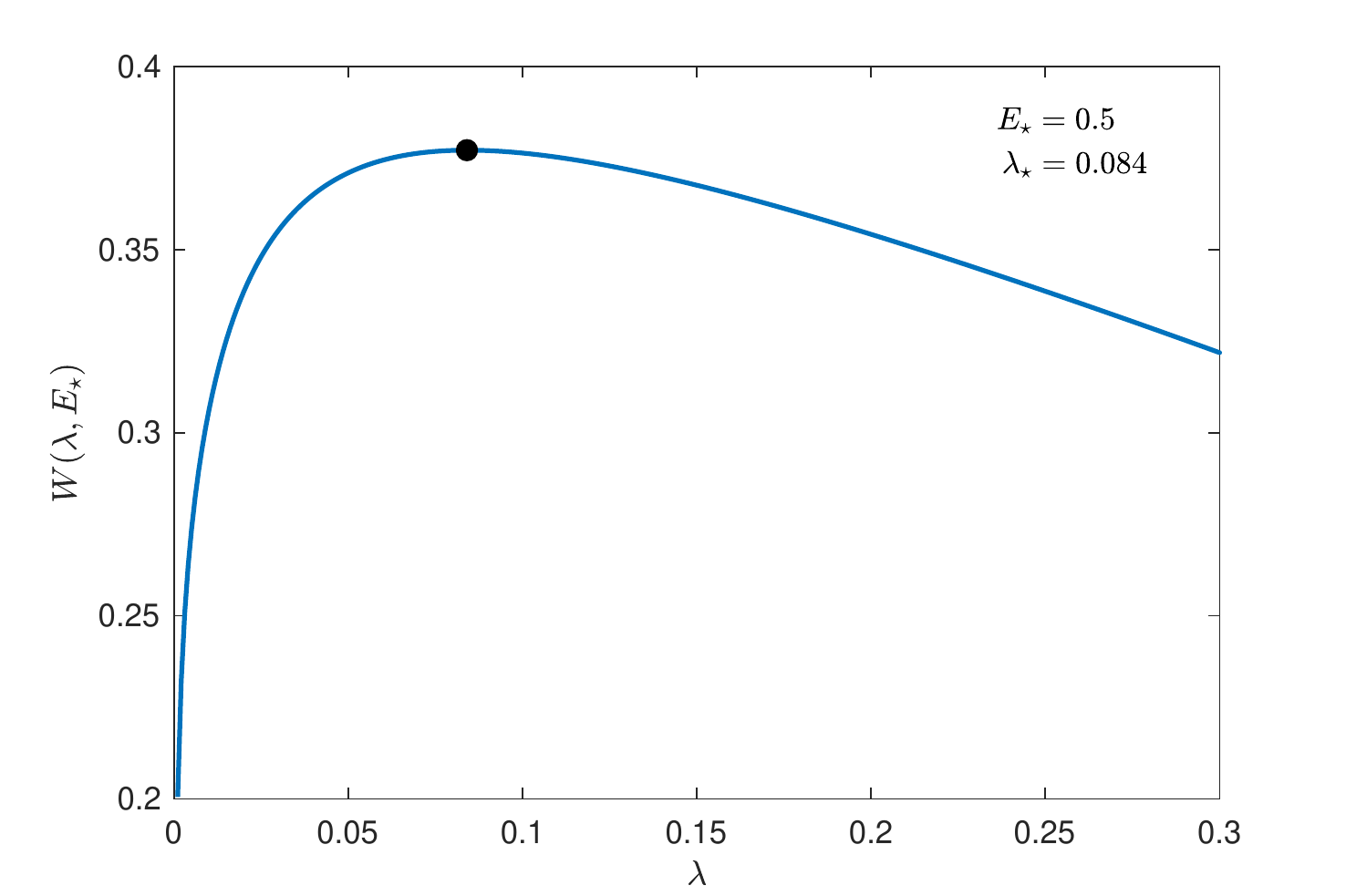}
\caption{\label{fig:wvslambda} 
The function $W(\lambda,E_\star)$ in the case of the lattice QCD correlator discussed in section~\ref{sec:lqcd} at $E_\star=0.5$. This function has a characteristic shape exhibiting a maximum at the optimal value $\lambda_\star$ of the trade--off parameter where the deterministic and error functionals are equally important in the minimization procedure.}
\end{figure}

The relative size of the statistical and systematic errors can be regulated by changing the parameter $\lambda$ and, if the estimate of the systematic uncertainty is reliable, results corresponding to different values of $\lambda$ have to be compatible within the total uncertainties. In our method the choice of the trade--off parameter can be optimized by using 
\begin{flalign}
W(\lambda,E_\star)=W[\lambda,g(\lambda,E_\star)]\;,
\end{flalign}
i.e. the function of $\lambda$ obtained by evaluating the functional $W[\lambda,g]$ at the solution $g_t(\lambda,E_\star)$ of the minimization procedure. This function has a characteristic shape that we show in Figure~\ref{fig:wvslambda}. At very small values $\lambda$ the contribution to $W$ coming from the error functional $\lambda B/C(0)^2$ is very small for generic values of the coefficients and the minimization procedure acts on the deterministic functional $(1-\lambda)A$ in order to obtain the best approximation of the target smearing function. Conversely, at very small values of $(1-\lambda)$ the contribution of the error functional $\lambda B/C(0)^2$ is dominant and the minimization procedure acts to reduce the statistical errors at the price of distorting the smearing function. The interplay between these two regimes generates a maximum in $W(\lambda,E_\star)$, 
\begin{flalign}
\max_\lambda\left\{ W(\lambda,E_\star)\right\}= W(\lambda_\star,E_\star)\;,
\end{flalign}
at the value $\lambda_\star$ where the deterministic and error functional are balanced. Therefore our method automatically suggests the optimal\footnote{We cannot exclude the presence of more than one maximum in $W(\lambda,E_\star)$. We never encountered this situation in our numerical experiments but, if this happens, we suggest to define $\lambda_\star$ as the smallest value of $\lambda$ where $W(\lambda,E_\star)$ has a maximum.} choice for the trade--off parameter, i.e. $\lambda=\lambda_\star$. 
    
In all our numerical experiments we have checked that the results corresponding to values of $\lambda$ smaller than $\lambda_\star$ are compatible within the corresponding total uncertainties. Indeed, statistical errors increase by decreasing $\lambda$ while the relative deviation $\delta_\sigma(E_\star,E)$ gets smaller and smaller and, in this region, Eq.~(\ref{eq:systerror}) can safely be used to get a reliable estimate of the systematic error. For values of $\lambda$ much larger than $\lambda_\star$, the results are instead affected by unacceptably large systematic uncertainties.

\section{The benchmark system}
\label{sec:benchmark}
In order to test our method and to compare it with the original Backus--Gilbert proposal we consider in this section a benchmark system where we know the exact spectral density. This information is used to build an exact synthetic correlator which we can then manually distort by adding random noise. We decided to consider the same benchmark system used in Ref.~\cite{Hansen:2017mnd} where additional details on the model can be be found.

The benchmark system is a toy model of three scalar particles, the pion $\pi$, the kaon $K$ and the $\phi$ meson with physical masses such that
\begin{flalign}
3m_\pi< 2m_K < m_\phi \;.
\end{flalign}
The particles are subject to the interaction Lagrangian density
\begin{flalign}
\mathcal{L}_{int}(x)= \frac{g_\pi}{6} \phi(x) \pi^3(x)+ \frac{g_K m_\phi}{2} \phi(x) K^2(x)
\end{flalign}
and the interactions are assumed to be perturbative. The authors of Ref.~\cite{Hansen:2017mnd} have considered a correlator in this theory having as finite volume spectral density the following expression,
\begin{flalign}
&\rho_L(E) = 
\frac{g_K^2 m_\phi^2}{2(m_\pi L)^3}\sum_{\vec p} \frac{\delta(E-2E_K(\vec p))}{4E_K^2(\vec p)}
\nonumber \\
&+
\frac{g_\pi^2}{48m_\pi^3 L^6}\sum_{\vec p,\vec q} \frac{\delta(E-E_\pi(\vec p)-E_\pi(\vec q)-E_\pi(\vec p+\vec q))}{E_\pi(\vec p)E_\pi(\vec q)E_\pi(\vec p+\vec q)}
\;,
\label{eq:rho_benchmark}
\end{flalign}
where the momenta are the ones allowed by periodic boundary conditions in space, i.e. $\vec p= 2\pi \mathbb{N}^3/L$, and where the energies are
\begin{flalign}
E^2_\pi(\vec p) = m_\pi^2+\vec p^2\;,
\quad
E^2_K(\vec p) = m_K^2+\vec p^2\;.
\end{flalign}
The infinite--volume spectral density is given by
\begin{flalign}
&\rho(E) = \frac{g_K^2 m_\phi^2}{32\pi^2 m_\pi^3}\sqrt{1-\frac{4m_K^2}{E^2}}\, \theta(E-2m_K)
\nonumber \\
&+
\frac{g_\pi^2}{3072\pi^4\, m_\pi}\left(\frac{E}{m_\pi}\right)^2\mathcal{F}\left(\frac{E}{m_\pi}\right)\,
\theta(E-3m_\pi)\;,
\label{eq:infrho}
\end{flalign}
where
\begin{flalign}
&\mathcal{F}(x)=
\nonumber \\
&
\frac{2}{x^4}\int_4^{(x-1)^2}dy \sqrt{
(y-4)\left[
\frac{(x^2-1)^2}{y}-2(x^2+1)+y
\right]
}\;.
\end{flalign}
The previous result agrees\footnote{Notice that our definition of $\rho(E)$ corresponds to $\rho(E)/2\pi$ in Ref.~\cite{Hansen:2017mnd}.} with the one originally given in Ref.~\cite{Hansen:2017mnd}.

In our numerical experiments we have set the parameters of the model to the same values used in Ref.~\cite{Hansen:2017mnd}, i.e. we have set $m_\pi=0.066$, $m_K/m_\pi=3.55$, $m_\phi/m_\pi=7.30$, $g_K=1$ and $g_\pi=10\sqrt{8}$. Since we are working in lattice units the previous numbers have to be read under the formal assumption that $a=1$. In evaluating the finite volume spectral density we have used a cutoff in the energy by replacing $\rho_L(E)$ with $\rho_L(E)\theta(\Lambda-E)$ using the value $\Lambda\approx 19m_\pi$.

\subsection{Exact data}
%
\begin{figure}[!t]
\includegraphics[width=0.48\textwidth]{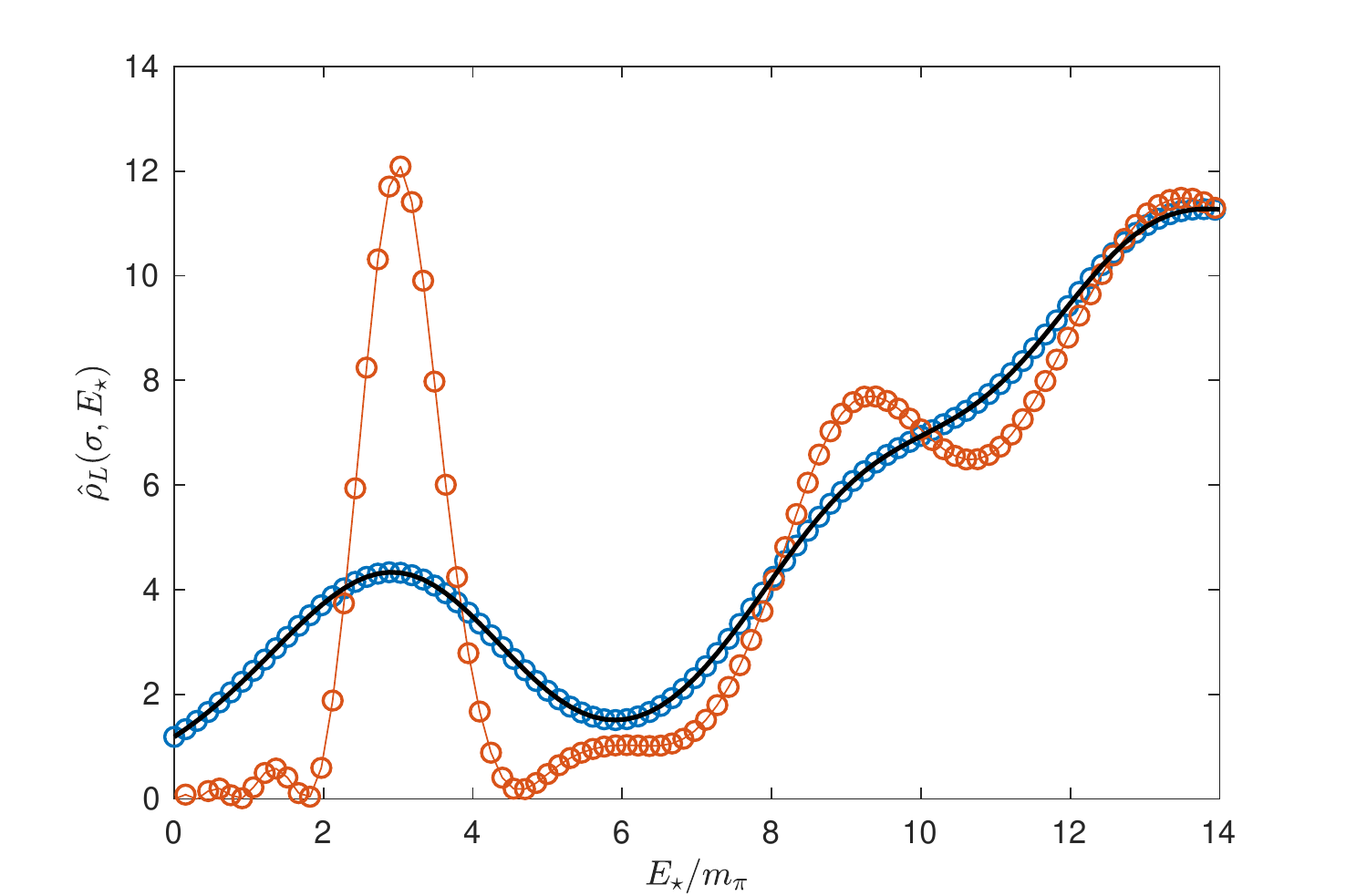}
\includegraphics[width=0.48\textwidth]{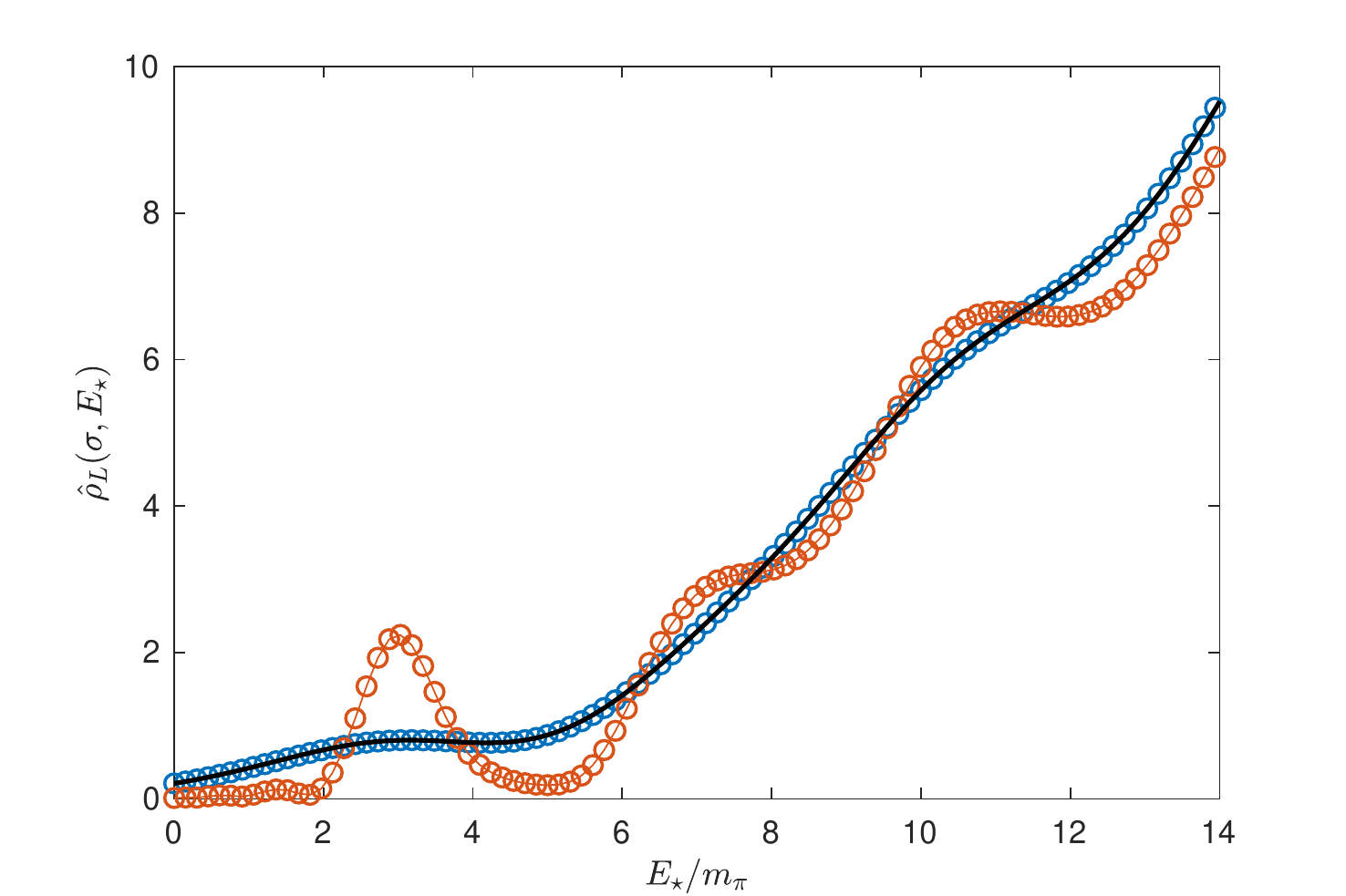}
\caption{\label{fig:noerr} 
Comparison of the results obtained with our method (blue points) with the results obtained by using the Backus--Gilbert method (orange points) in the absence of statistical errors. The two plots correspond to the volumes $L=24$ (first row) and $L=32$ (second row) with $t_{max}=30$ in both cases. In the case of our method the results have been obtained by setting $E_0=0$ and by using the Gaussian of Eq.~(\ref{eq:gaussian}) as target smearing function with $\sigma=0.1$. In both plots the solid black curve corresponds to the exact smeared spectral density that our method (not the Backus--Gilbert one) is expected to reproduce in the infinite $t_{max}$ limit.}
\end{figure}
In Figure~\ref{fig:noerr} we compare the results obtained with our method (blue points) in the absence of statistical errors with the results obtained by using the Backus--Gilbert method (orange points). The two plots in the figure have been obtained by starting from the correlator
\begin{flalign}
C(t)=\int_0^{\infty} dE\, \rho_L(E)\, b_{\infty}(t,E)
\end{flalign}
on the volumes $L=24$ (first row) and $L=32$ (second row) with $t_{max}=30$. In the case of our method the results have been obtained by setting $E_0=0$ and by using the target smearing function in Eq.~(\ref{eq:gaussian}) with $\sigma=0.1$. In both plots the solid black curve corresponds to the exact smeared spectral density,
\begin{flalign}
\hat \rho_L(\sigma,E_\star)=\int_0^{\infty} dE\, \rho_L(E)\, \Delta_\sigma(E_\star,E)\;,
\end{flalign}
which our method is expected to reproduce in the infinite $t_{max}$ limit. As it can be seen, in both plots the agreement between the numerical results obtained with our method and the exact result is excellent. Notice that the results obtained with the Backus--Gilbert method are not expected to reproduce the black line. In fact, because the smearing function is an output of the procedure, it can only be controlled by changing $t_{max}$ and, moreover, it is different at different values of $E_\star$. We can only notice that our choice of setting $\sigma=0.1$ is similar to the choice made by the Backus--Gilbert method on the volume $L=32$ at high energies where the smeared spectral density is more smooth and starts to have an infinite--volume like behaviour.  

Additional examples of applications of our method in the case of exact synthetic data can be found in Appendix~\ref{sec:appendix2}.

\subsection{Noisy data}
In order to test our method in the case of a noisy correlator, we add uncorrelated random noise to the synthetic correlator in such a way that the signal-to-noise ratio is constant. For the results presented here, on average, the relative standard deviation of the correlator is chosen to be around 2\%. For all the numerical examples shown in the rest of the paper, we only use the diagonal part of the covariance matrix in the minimization procedure.

In the reconstruction of the spectral density we estimate the statistical and systematic uncertainty independently and combine them in quadrature. To estimate the statistical uncertainty we use a bootstrapping procedure, i.e.~we apply our method to a set of bootstrap samples, from which we derive the mean and standard deviation of the reconstructed spectral density. For the systematic uncertainty we use Eq.~\eqref{eq:systerror} such that the total uncertainty is given by
\begin{equation}
 \Delta^{total} = \sqrt{(\Delta^{stat})^2+(0.68\times\Delta^{syst})^2}~,
\end{equation}
where the factor $0.68$ is introduced to give a consistent $1\sigma$ uncertainty on the final result.

\begin{figure}[!t]
\includegraphics[width=0.48\textwidth]{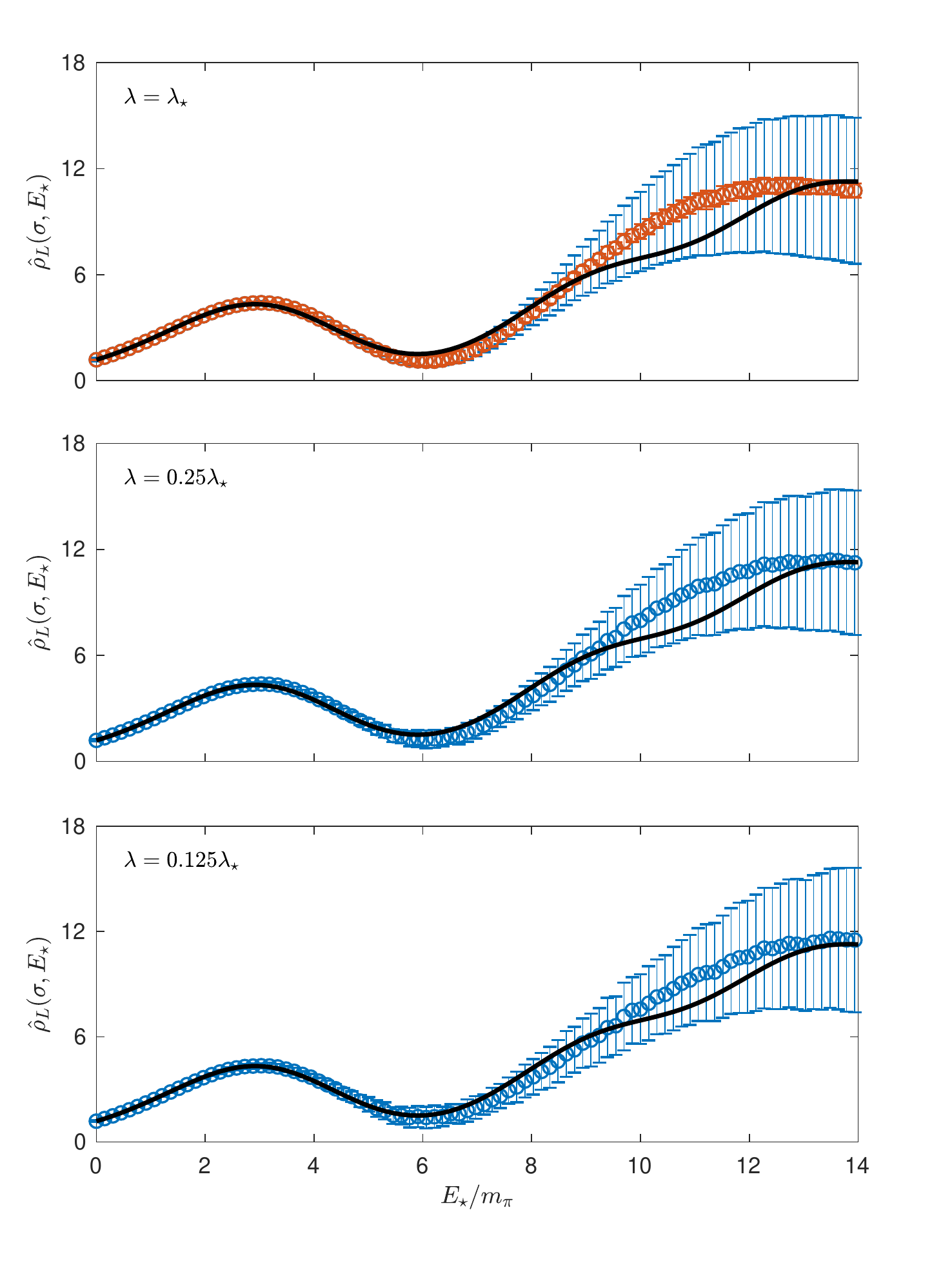}
\caption{\label{fig:lambda_dep} 
Examples of how the reconstructed spectral density depends on the $\lambda$ parameter. On the topmost plot, the orange band shows the statistical uncertainty, while in all cases the blue band shows the combined uncertainty (statistical and systematic). For this particular case, where $L=24$ and $t_{max}=30$ with $\sigma=0.1$,  the dependence on $\lambda$ is practically negligible. These results have been obtained by using $b_\infty(t,E)$ as basis functions.}
\end{figure}
In Figure~\ref{fig:lambda_dep} we show three plots in order to discuss the impact of the choice of the trade--off parameter $\lambda$ and the significance of our estimate of the systematic error. In all the plots the black curve is the exact smeared spectral density. The data in the top--panel have been obtained at the value of $\lambda_\star$ determined from the maximum of $W(\lambda,E_\star)$ at $E_\star/m_\pi=7$ and, after having checked that in this case $\lambda_\star$ is not strongly dependent upon $E_\star$, we have used the same value of the trade--off parameter at all energies. On the first plot, the orange band shows the statistical error, while in all plots, the blue band corresponds to the total uncertainty. As it can be seen, the data are in very good agreement with the exact result already at the $1\sigma$ level of uncertainty. The plots in the center and bottom panels have been obtained by using respectively $\lambda=\lambda_\star/4$ and $\lambda=\lambda_\star/8$ at all energies. For this particular case, where $L=24$ and $t_{max}=30$ with $\sigma=0.1$,  the dependence on $\lambda$ is practically negligible. Similar results can be shown at different volumes, at different values of $t_{max}$ and at different values of $\sigma$. The results shown in rest of this section have all been obtained at the value of $\lambda_\star$ determined from the maximum of $W(\lambda,E_\star)$ at $E_\star/m_\pi=7$ for all energies.

\begin{figure}[!t]
\includegraphics[width=0.48\textwidth]{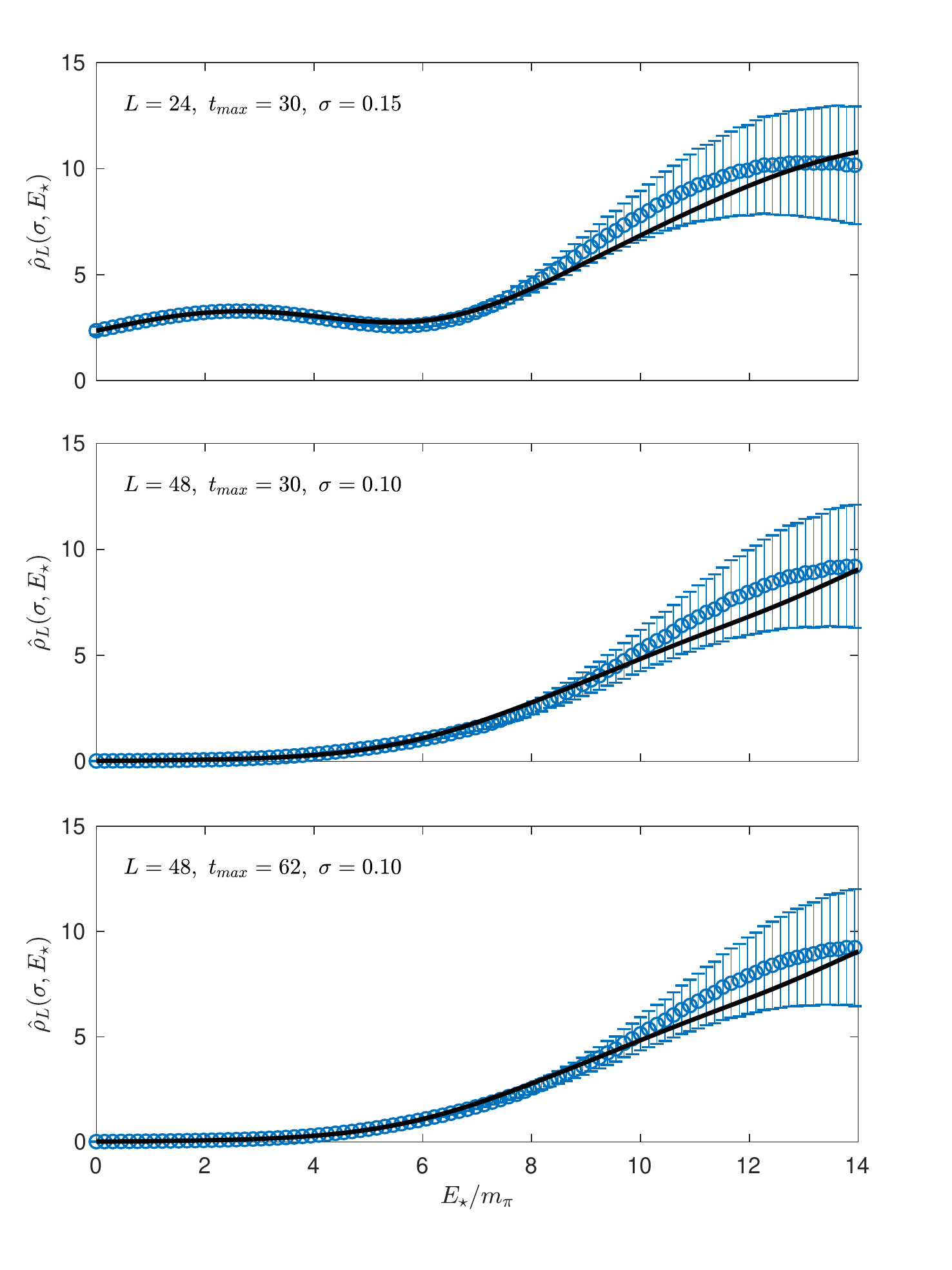}
\caption{\label{fig:smoother} 
In cases where the smeared spectral density is very smooth, either because the smearing parameter $\sigma$ is larger (topmost plot), or because we are closer to the infinite volume result (second and third plot), the quality of the reconstruction is much better. For the second and third plot we use two different values of $t_{max}$, and while increasing $t_{max}$ does improve the result slightly, already at $t_{max}=30$ we have full control over the reconstruction of the spectral density. These results have been obtained by using $b_\infty(t,E)$ as basis functions.}
\end{figure}
The results shown in Figure~\ref{fig:lambda_dep} correspond to a relatively challenging situation because on small volumes and/or at small values of $\sigma$ the smeared spectral function exhibits an oscillating behaviour induced by the fact that the energy levels are largely spaced (see also Appendix~\ref{sec:appendix2} for other examples). In cases where the smeared spectral density is very smooth, either because the smearing parameter $\sigma$ is larger, or because the volume is larger, the quality of the reconstruction is much better. This can be seen in Figure~\ref{fig:smoother} where we show results on the volume $L=24$ corresponding to a larger value of $\sigma$ with respect to the one used in Figure~\ref{fig:lambda_dep} and results on the volume $L=48$ for two different values of $t_{max}$. In this cases the use of Eq.~\eqref{eq:systerror} to quantify the systematic uncertainty results in an over--estimate of the error. This feature makes us pretty confident about the reliability of the results obtained with our method.

\begin{figure}[!t]
\includegraphics[width=0.48\textwidth]{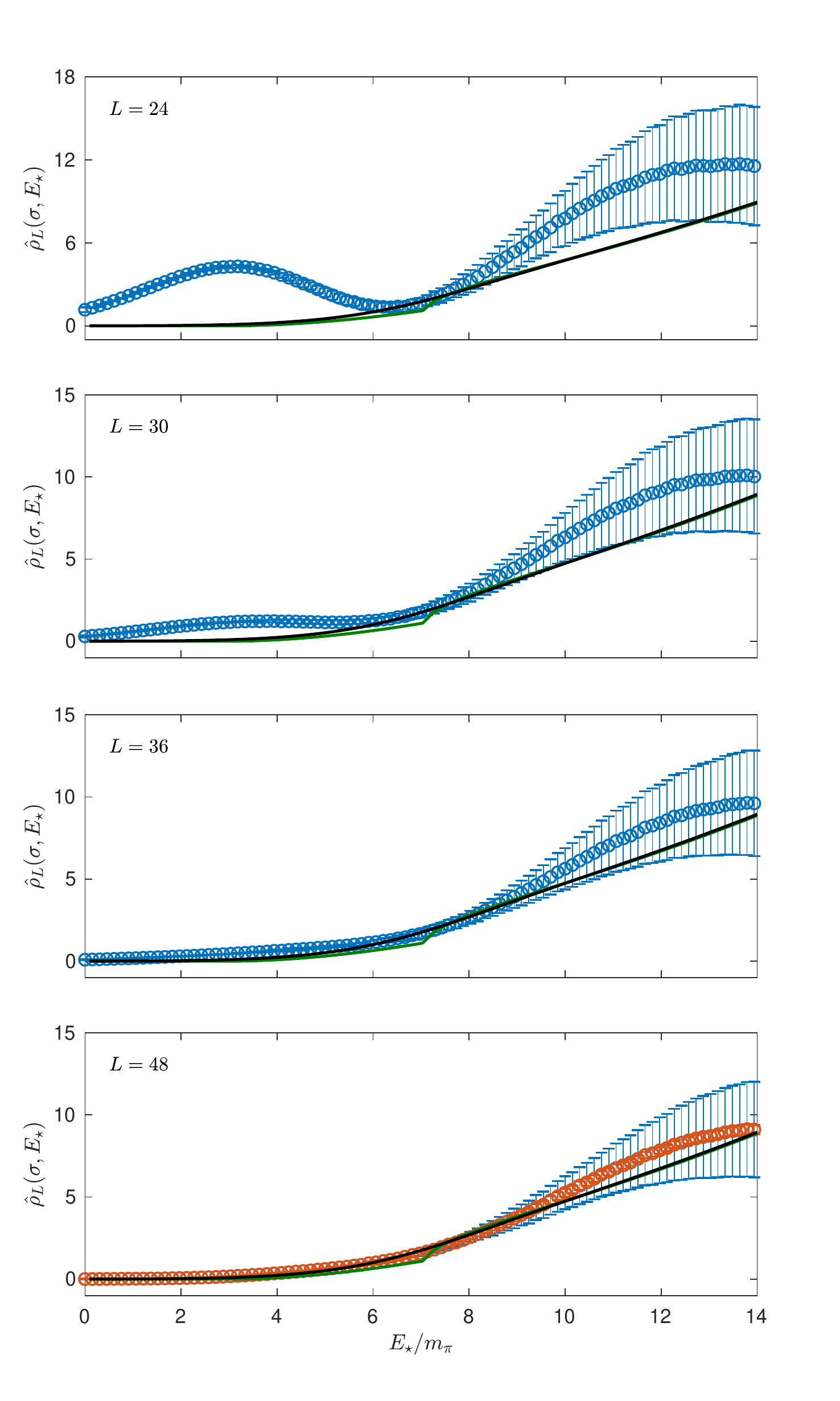}
\caption{\label{fig:infvol} 
Approach to the infinite--volume limit of the reconstructed smeared spectral densities. The results correspond to $\sigma=0.1$ and have been obtained by using $b_{T}(t,E)$ as basis functions with $E_0=0$, $T=2(t_{max}+1)$ and $t_{max}=31$. Starting from the top, the data in the first panel correspond to $L=24$, those in the second panel to $L=30$, those in the third panel to $L=36$ and those in the last panel to $L=48$. In all plots the green and black curves correspond respectively to the exact infinite--volume unsmeared and smeared spectral densities. The orange band in the last plot corresponds to the statistical uncertainties. The numerical data have to reproduce the black curve in the infinite--volume limit and, within the errors, the agreement is already very good at $L=36$. Remarkably, at $L=48$ the numerical data agree with the infinite volume curve up to energies of order $E_\star/m_\pi\simeq 11$ at the level of the statistical uncertainties. 
}
\end{figure}

We close this section by illustrating the approach to the infinite--volume limit of the reconstructed smeared spectral densities in the case of this benchmark model. The plots in Figure~\ref{fig:infvol} show the results obtained on different volumes by setting the smearing radius parameter to $\sigma=0.1$ and by using $b_{T}(t,E)$ as basis functions with $T=2(t_{max}+1)$ and $t_{max}=31$ (see Appendix~\ref{sec:appendix2} for another example of this analysis). More precisely, the plot in the first panel (starting from the top) corresponds to $L=24$, the one in the second panel to $L=30$, the one in the third panel to $L=36$ and the one in the last panel to $L=48$. In all plots the green curve corresponds to $\rho(E)$ of Eq.~(\ref{eq:infrho}) while the black curve corresponds to the exact smeared infinite--volume spectral density, namely
\begin{flalign}
\hat \rho(\sigma,E_\star)=\int_0^{\infty} dE\, \rho(E)\, \Delta_\sigma(E_\star,E)\;.
\end{flalign}
As it can be seen, $\rho(E)$ is a continuous function of the energy for $E\ge 3m_\pi$ but it has a cusp at $E=2m_K$, i.e. in correspondence of the two--kaons threshold. In the infinite--volume limit the numerical data are expected to reproduce  the \emph{smeared} spectral density that is instead a smooth function. This already happens, within the errors, at $L=36$ for medium--high values of the energy and for all the explored energies at $L=48$. Remarkably, at $L=48$ the numerical data agree with the infinite volume curve up to energies of order $E_\star/m_\pi\simeq 11$ at the level of the statistical uncertainties (orange band) thus confirming that on large volumes Eq.~(\ref{eq:systerror}) gives a conservative estimate of the systematic uncertainties.

\section{Lattice correlators}
\label{sec:lqcd}
%
\begin{figure}[!t]
\includegraphics[width=0.48\textwidth]{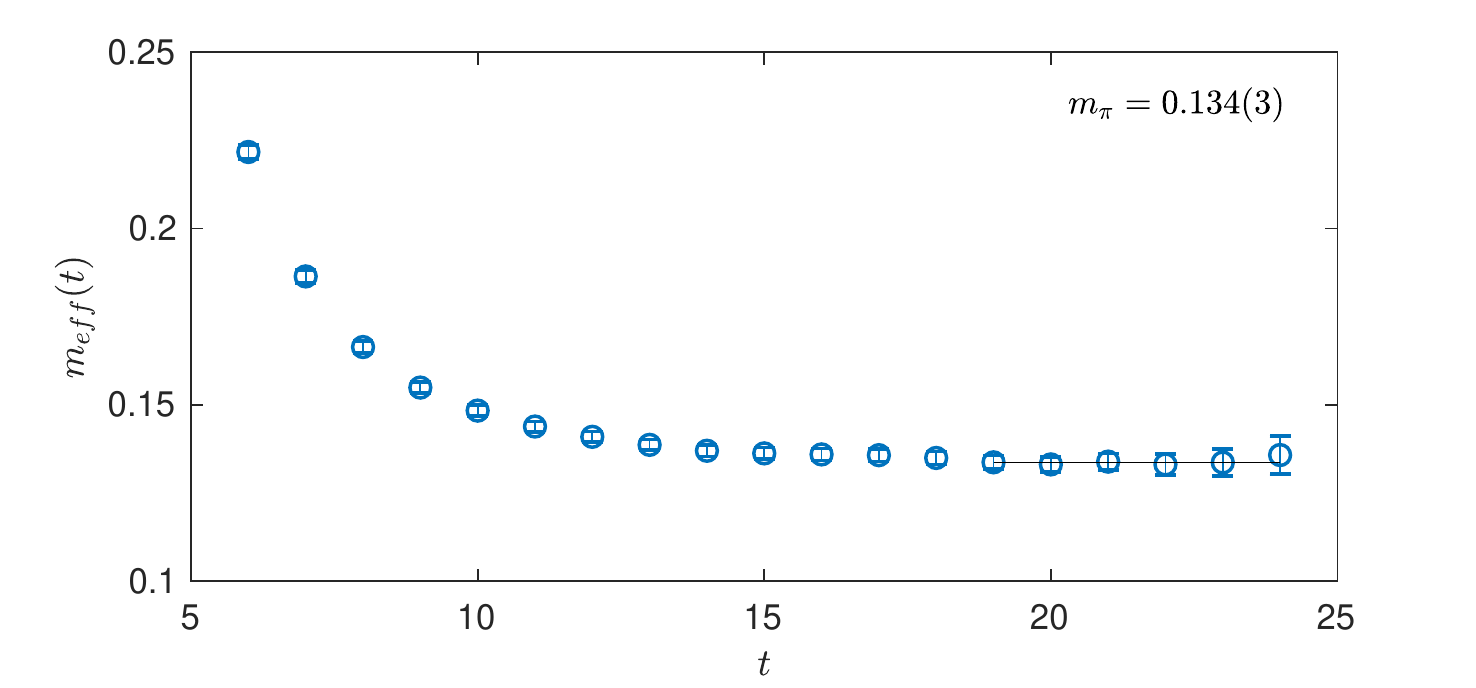}
\includegraphics[width=0.48\textwidth]{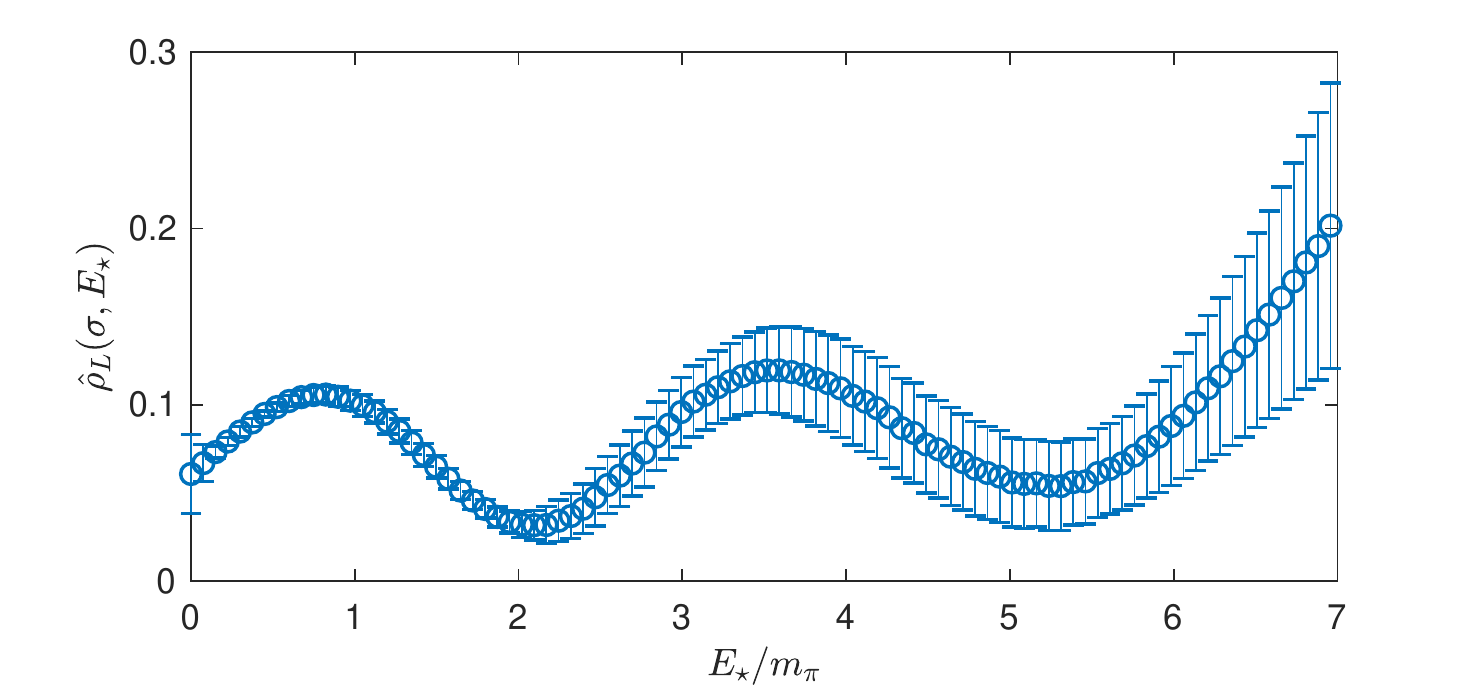}
\caption{\label{fig:b2} 
Extraction of the smeared spectral density from the correlator $C_\text{QCD}(t)$ discussed in the text. The top--panel shows the calculation of the pion mass, the lightest state contributing to the spectral density in this case, extracted from a standard effective--mass analysis. The bottom--panel shows the reconstructed smeared spectral density obtained by applying our method with $\sigma=0.1$, by using $b_T(t,E)$ as basis functions with $T=48=(2t_{max}+1)$, by setting $E_0=0.37m_\pi$ and by using the value of $\lambda_\star$ determined at $E_\star=3.7 m_\pi$ for all the energies explored. As expected, the smeared spectral density shows a peak in correspondence of $E_\star/m_\pi\simeq 1$ and another structure around $E_\star/m_\pi\simeq 3$.
}
\end{figure}
In this section, in order to show the quality of the results that can be obtained by applying our method to true data, we discuss two analyses of simulated lattice correlators from which we extract the associated smeared spectral densities.

In the first example we consider a meson pseudoscalar--pseudoscalar correlator obtained by performing a lattice simulation of QCD with three degenerate flavours on a lattice volume $L^3\times T=24^3\times 48$ with periodic boundary conditions in time and $C^*$ boundary conditions~\cite{Lucini:2015hfa} along the spatial directions. 
The bare parameters of the simulation correspond to the CLS ensemble H101 and can be found in Table~1 of Ref.~\cite{Bruno:2016plf}.  
More precisely, the correlator is given by
\begin{flalign}
&
C_\text{QCD}(t)=\frac{1}{2L^3}\sum_{\vec x}\, T\bra{0}\, P(0)\,  P(x)\, \ket{0}\;,
\nonumber \\
&
P(x)=
\left\{ \bar d \gamma_5 u+  \bar u \gamma_5 d\right\}(x)\;,  
\end{flalign}
where $u$ and $d$ are the up and down quark fields that, in this unphysical simulation, have the same mass. The lightest states contributing to the finite volume spectral density associated with the correlator $C_\text{QCD}(t)$ are expected to be the pion and the three--pions states with vanishing total momentum allowed by the boundary conditions. This means that we expect the leading contributions to $\rho_L(E)$ to be proportional to $\delta(E-m_\pi)$ and to $\delta(E-E_{3\pi})$, where $m_\pi$ is the mass of the pion and $E_{3\pi}\simeq 3m_\pi$. In the top--panel of Figure~\ref{fig:b2} we show the numerical determination of $m_\pi$ from a standard effective--mass analysis. In the bottom panel of the same figure we show the smeared spectral density obtained by applying our method with $\sigma=0.1$ and by using the value of $\lambda_\star$ determined at $E_\star=0.5\simeq 3.7 m_\pi$ (see Figure~\ref{fig:wvslambda}) for all the energies explored. As it can be seen, the reconstructed smeared spectral density clearly shows a peak centred around $E_\star/m_\pi\simeq 1$ and another structure around $E_\star/m_\pi\simeq 3$. 

\begin{figure}[!t]
\includegraphics[width=0.48\textwidth]{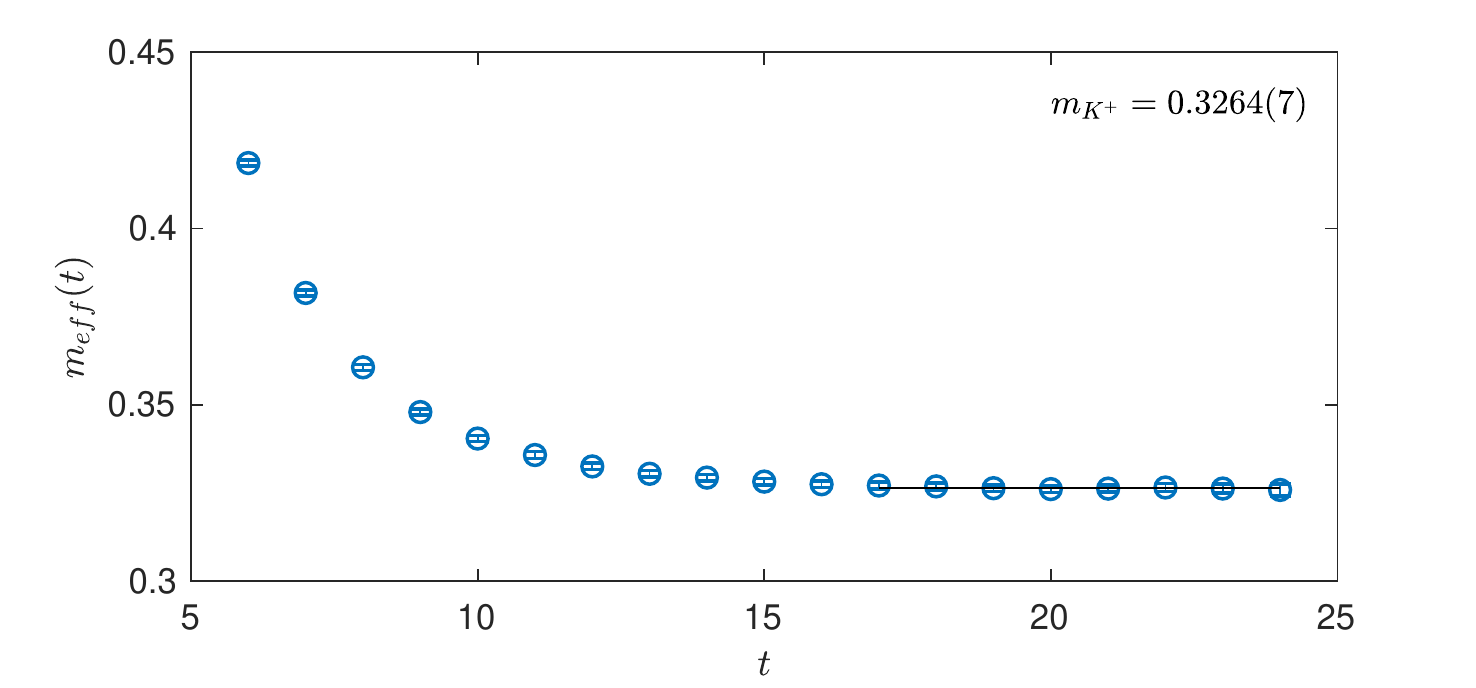}
\includegraphics[width=0.48\textwidth]{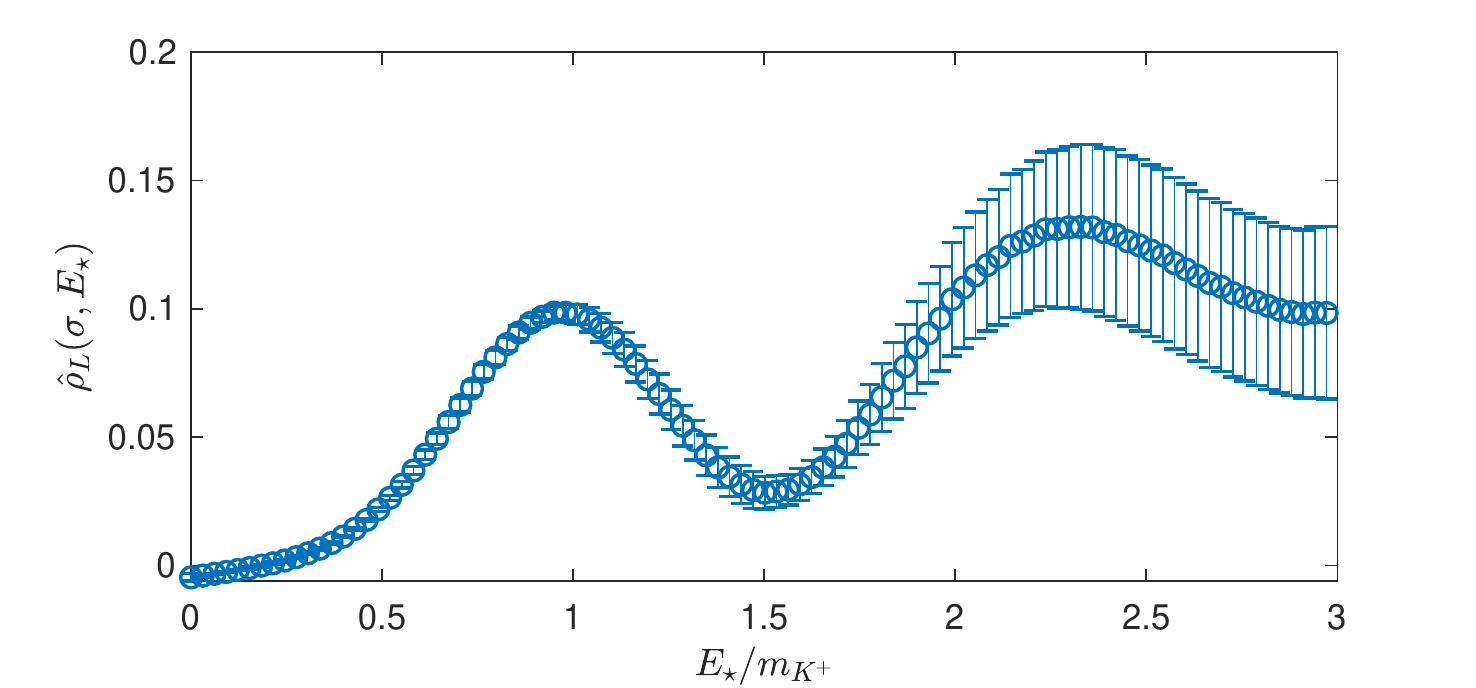}
\caption{\label{fig:q3} 
Extraction of the smeared spectral density from the correlator $C_{\text{QCD}+\text{QED}}(t)$ discussed in the text. The top--panel shows the calculation of the charged kaon mass, the lightest state contributing to the spectral density in this case, extracted from a standard effective--mass analysis. The bottom--panel shows the reconstructed smeared spectral density obtained by applying our method with $\sigma=0.1$, by using $b_T(t,E)$ as basis functions with $T=48=(2t_{max}+1)$, by setting $E_0=0.15m_{K^+}$ and by using the value of $\lambda_\star$ determined at $E_\star=1.5 m_{K^+}$ for all the energies explored. As expected, the smeared spectral density shows an isolated peak in correspondence of $E_\star/m_{K^+}\simeq 1$ and another structure that starts in proximity of $E_\star/m_{K^+}\simeq 2.4$.
}
\end{figure}
In the second example we consider again a meson pseudoscalar--pseudoscalar correlator, but in this case it has been obtained from a QCD$+$QED simulation performed at the unphysical value $\alpha_{em}=0.05$ of the electromagnetic coupling constant with dynamical up, down and strange quarks. The masses of the down and the strange quarks, having the same negative electric charge, have been set equal in this simulation and different from the mass of the positively charged up quark. The simulation has been performed on a lattice volume $L^3\times T=24^3\times 48$ with periodic boundary conditions in time and $C^*$ boundary conditions~\cite{Lucini:2015hfa} along the spatial directions. More details about the simulation can be found in Ref.~\cite{Hansen:2018zre}. The plot in the top--panel of Figure~\ref{fig:q3} shows the effective mass extracted from the correlator
\begin{flalign}
&
C_{\text{QCD}+\text{QED}}(t)=\frac{1}{2L^3}\sum_{\vec x}\, T\bra{0}\, 
P(0)\, P(x)\, \ket{0}\;,
\nonumber \\
&
P(x)=
\left\{ \bar S \gamma_5 U+  \bar U \gamma_5 S\right\}(x)\;,  
\end{flalign}
where $S(x)$ and $U(x)$ are the gauge--invariant interpolating operators for the strange and up quarks described in details in Refs.~\cite{Lucini:2015hfa,Hansen:2018zre}. In this case the effective mass analysis measures $m_{K^+}$, the mass of the charged kaon, and this is the lightest state contributing to the finite volume spectral density. Increasing the energy, we expect contributions to the spectral density coming from states corresponding to the charged kaon plus photons and from states with three kaons. Since the volume is rather small in this case and the boundary conditions do not allow for the propagation of photons with zero momenta, after the charged kaon peak we expect a contribution to $\rho_L(E)$ proportional to $\delta(E-E_{3K})$ with $E_{3K}\simeq m_{K^+}+2m_{K^0}$. By using the value of $m_{K^0}$ measured in Ref.~\cite{Hansen:2018zre} we have $E_{3K}/m_{K^+}\simeq 2.6$. This expectation is confirmed by the plot in the bottom--panel of Figure~\ref{fig:q3} where the smeared spectral density shows an isolated peak in correspondence of $E_\star/m_{K^+}\simeq 1$ and a structure that starts in proximity of $E_\star/m_{K^+}\simeq 2.4$.

\section{Conclusions}
\label{sec:conclusions}
We have presented a new method for addressing inverse problems in the presence of noisy observations. The method can be used to extract smeared spectral densities from measured correlation functions and it provides results associated with a reliable estimate of both the statistical and the systematic uncertainties.

The function used for smearing the spectral density is an input of our method, and for this reason, it can be held fixed in the analysis of data corresponding to different correlators. This feature is particularly convenient in lattice applications because it allows to study the infinite volume limit of the reconstructed smeared spectral densities in a systematic way. 

The mechanism used in our method to keep statistical errors under control has been inherited from the classical Backus and Gilbert approach. The method has a natural built--in mechanism to optimize the choice of the so--called trade--off parameter and, moreover, the significance of the estimate of the errors can be assessed by checking the compatibility of the results obtained at sub--optimal values of this parameter.      

In order to illustrate the quality of the results that can be obtained with our method, we have applied it to a benchmark system where we know the exact spectral density, both on finite volumes and in the infinite volume limit. We have shown that the results obtained with our approach reproduce within the errors the expected finite volume smeared spectral densities and also that, by increasing the volume, they approach the expected infinite volume limit. 

We have also applied the method to true data in the case of correlators obtained from QCD and QCD$+$QED lattice simulations. Using these examples we have shown that smeared spectral densities can be extracted with satisfactory accuracy and that the numerical results are compatible with the expectations coming from the knowledge of the spectrum of the two theories. 

We have discussed the method by using the language of lattice correlators but, given its generality, we are pretty confident that, together with other valuable approaches already present in the literature (see for example the already quoted Refs.~\cite{Asakawa:2000tr,Burnier:2013nla,Rothkopf:2016luz}), it will be useful to address inverse problems arising in other fields of research, particularly those where the classical Backus--Gilbert method has already proven useful. 

\begin{acknowledgments}
Useful conversations with A.~Patella are gratefully acknowledged. N.T. thanks the University of Rome Tor Vergata for partial support under the project PLNUGAMMA.
\end{acknowledgments}

\appendix

\section{}
\label{sec:appendix}
%
\begin{figure}[!t]
\includegraphics[width=0.48\textwidth]{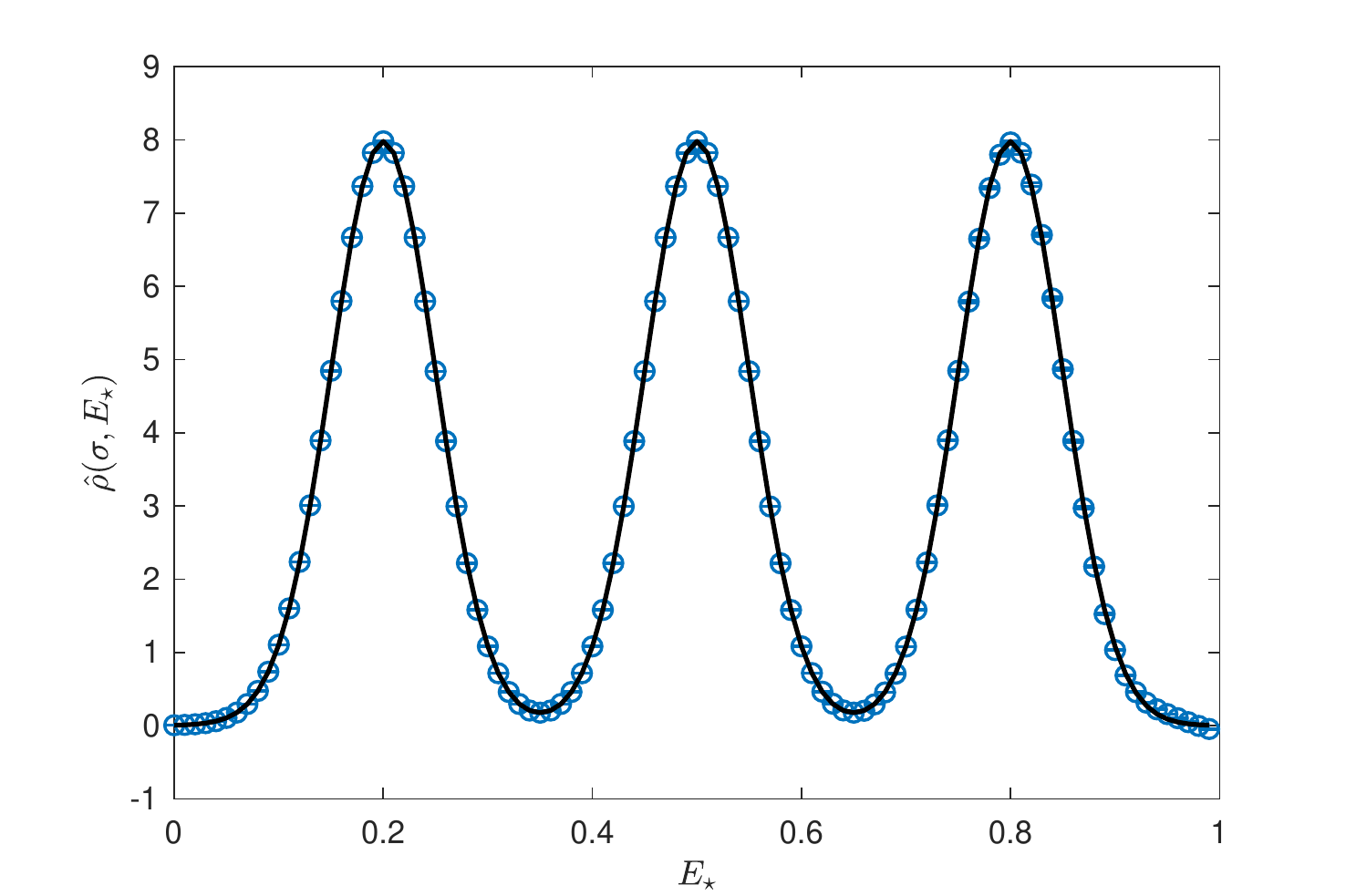}
\includegraphics[width=0.48\textwidth]{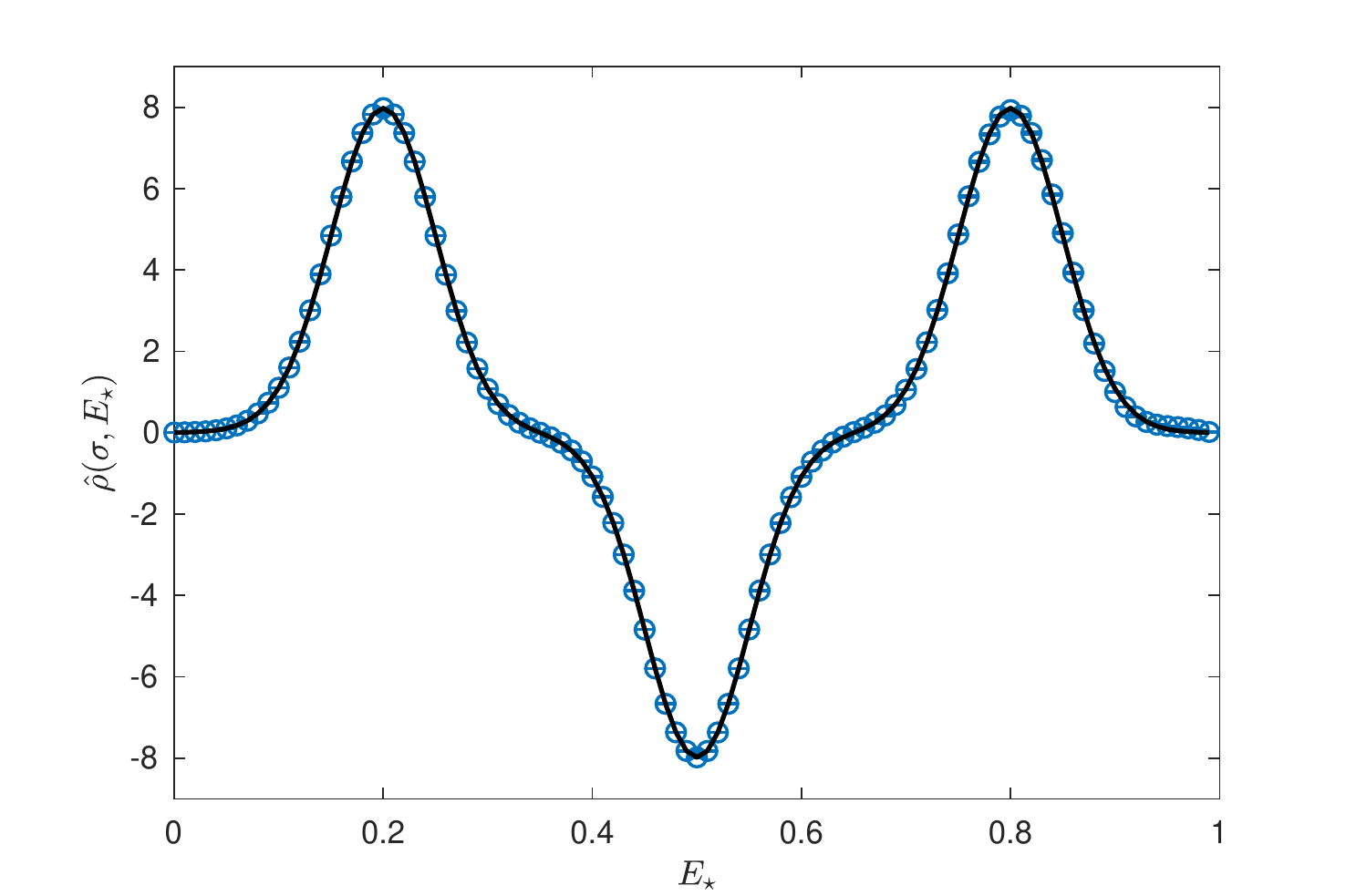}
\caption{\label{fig:ppp_pnp}
Example showing that our method is able to reconstruct the smeared spectral density from Eq.~\eqref{eq:ppp_pnp} regardless of the sign of the peaks. For the plot we use $b_\infty(t,E)$ as the basis functions with $t_{max}=62$ and the smallish value $\sigma=0.05$ to properly separate the peaks.
}
\end{figure}
In this short appendix we collect the explicit expressions for the formulas used in the numerical implementation of the presented method. The smearing function in Eq.~\eqref{eq:gaussian} can be written as
\begin{equation}
 \Delta_\sigma(E_\star,E) = \frac{1}{\sqrt{2\pi}\sigma Z}\exp\left(\frac{-(E-E_\star)^2}{2\sigma^2}\right),
\end{equation}
with the additional normalization factor
\begin{equation}
 Z = \frac{1}{2}\left(1+\mathrm{erf}\left(\frac{E_\star}{\sqrt{2}\sigma}\right)\right).
\end{equation}
In the numerical implementation, the functional $A[g]$ is defined as
\begin{equation}
 A[g] = \int_{E_0}^\infty dE~e^{\alpha E}\left\{\bar \Delta_\sigma(E_\star,E)-\Delta_\sigma(E_\star,E)\right\}^2.
\end{equation}
With this definition, the matrix $\mathtt{A}_{tr}$ is given by
\begin{align}
\begin{split}
 \mathtt{A}_{tr} &= \frac{e^{-(r+t+2-\alpha)E_0}}{r+t+2-\alpha}
 + \frac{e^{-(T-r+t-\alpha)E_0}}{T-r+t-\alpha} \\[2mm]
 &+ \frac{e^{-(T+r-t-\alpha)E_0}}{T+r-t-\alpha}
 + \frac{e^{-(2T-r-t-2-\alpha)E_0}}{2T-r-t-2-\alpha}~.
\end{split}
\end{align}
The parameter $\alpha$ allows for changing the measure as a function of the energy and it must satisfy $\alpha<2$. For all results presented in this paper we simply used $\alpha=0$. In the limit $T\to\infty$ only the first term contributes to the matrix. The vector $f_t$ is similarly defined as
\begin{equation}
 f_t = (1-\lambda)\int_{E_0}^\infty dE~e^{\alpha E}~\Delta(E,E_\star,\sigma)~b_T(t+1,E)~,
\end{equation}
whose components can be calculated as
\begin{equation}
 f_t = N(t+1)F(t+1) + N(T-t-1)F(T-t-1)~,
\end{equation}
using the auxiliary functions
\begin{align}
 N(k) &= \frac{1-\lambda}{2Z}\exp\left(\frac{(\alpha-k)((\alpha-k)\sigma^2+2E_\star)}{2}\right), \\[2mm]
 F(k) &= 1+\mathrm{erf}\left(\frac{(\alpha-k)\sigma^2+E_\star-E_0}{\sqrt{2}\sigma}\right).
\end{align}
Again, in the limit $T\to\infty$ only the first term contributes to the vector $f_t$ because $F(\infty)=0$. Finally we define the vector $R_t$ as
\begin{equation}
 R_t = \int_0^\infty dE~b_T(t+1,E) = \frac{1}{t+1}+\frac{1}{T-t-1}~,
\end{equation}
where once again the second term vanishes for $T\to\infty$.

\section{}
\label{sec:appendix2}
\begin{figure}[!t]
\includegraphics[width=0.48\textwidth]{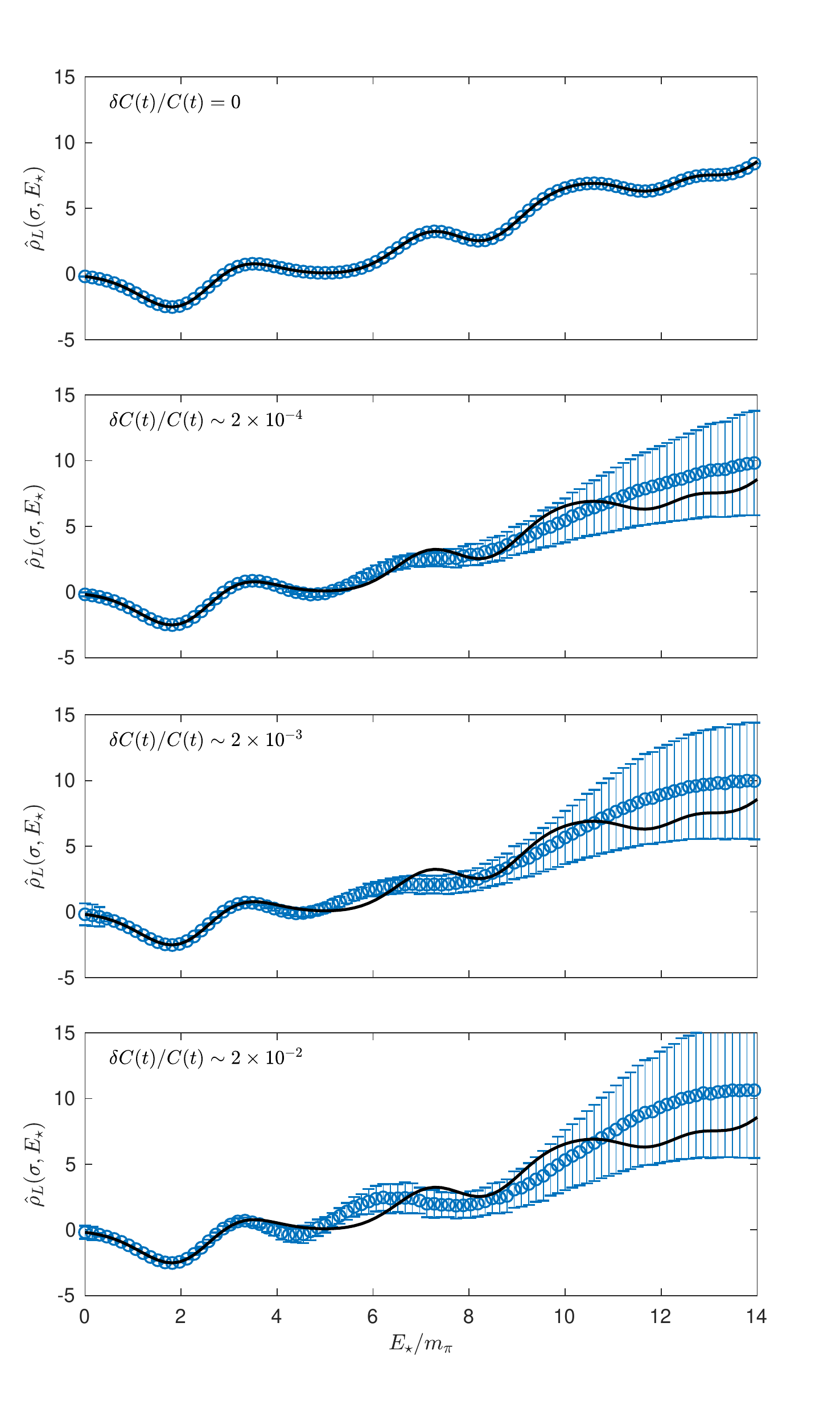}
\caption{\label{fig:2pion_neg}
Realistic example of a spectral density with a negative contribution at low energy, as described by Eq.~\eqref{eq:rho_neg}. Because our method has no constraints regarding the sign of the spectral density, we also observe a good reconstruction at low energy even when the spectral density is negative. The plot has been produced by using $b_\infty(t,E)$ as the basis functions with $t_{max}=126$ and $\sigma=0.05$.
}
\end{figure}
\begin{figure}[!t]
\includegraphics[width=0.48\textwidth]{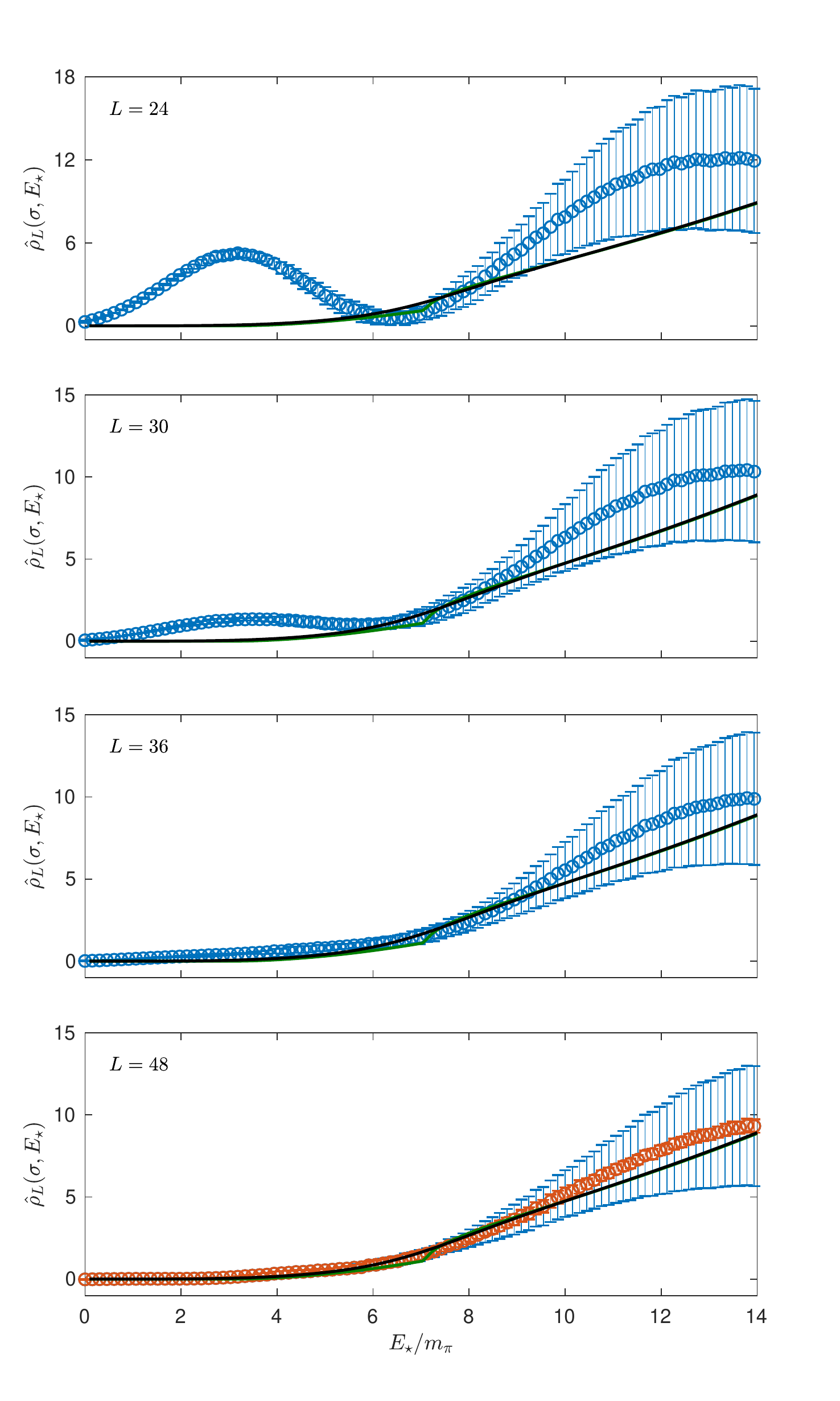}
\caption{\label{fig:infvol_075}
Reproduction of Figure~\ref{fig:infvol} using the smaller value $\sigma=0.075$ of the smearing parameter. As expected, in the first row the peak becomes sharper, and in all cases the error increases at high energy due to larger systematic uncertainties. 
}
\end{figure}

In this appendix, in order to highlight some features of our method, we present some more examples of reconstruction of synthetic spectral densities.

Our method does not require the knowledge of any prior information on the spectral density and, therefore, is a totally model--independent approach.
In particular, all spectral densities considered in the main text are non-negative but, in fact, our method is oblivious to the sign of the spectral density. In order to highlight this property we have considered an artificial finite--volume spectral density consisting of three separated peaks
\begin{equation}
 \rho(E) = \delta(E-0.2) \pm \delta(E-0.5)+\delta(E-0.8)~,
 \label{eq:ppp_pnp}
\end{equation}
where the second peak is either positive or negative. In Figure~\ref{fig:ppp_pnp} we show the exact smeared spectral density and the reconstruction, in the absence of statistical errors, for both choices of sign. For the plot we use $t_{max}=62$ and $\sigma=0.05$ and we observe that the reconstruction is equally good, regardless of the sign of the second peak.

As a more realistic example we consider the spectral density from Eq.~\eqref{eq:rho_benchmark} and add a new bound state with an energy $E=2m_\pi$ such that the spectral density reads
\begin{equation}
 \rho_L(E) \to \rho_L(E) - \frac{\delta(E-2m_\pi)}{40m_\pi}~.
 \label{eq:rho_neg}
\end{equation}
Because of the negative weight for this state, the smeared spectral density will be negative at low energy, as shown on Figure~\ref{fig:2pion_neg}. Starting from the top, the different rows of the figure represent the result obtained by assigning an increasing relative error to the correlator.  For the plots we use
\begin{equation*}
 \sigma = 0.05~, \quad t_{max} = 126~, \quad E_0 = 0.05~,
\end{equation*}
and we observe that the reconstruction is excellent at low energy and it captures the average behaviour at high energy.

Throughout most of the paper we used $\sigma=0.1$ as a reasonable choice of smearing parameter, a choice that is supported by the scan in Figure~\ref{fig:sigma_min}. To prove that our method also works for other (reasonable) choices of the smearing parameter, we reproduce Figure~\ref{fig:infvol} using the smaller value $\sigma=0.075$. This result is shown in Figure~\ref{fig:infvol_075} where, as expected, the peak at low energy and small volume becomes sharper, and the errors at high energy increase due to larger systematic uncertainties. However, the overall quality of the reconstruction is mostly unchanged.

Redoing this figure with a significantly smaller value of $\sigma$ requires a substantial decrease of the statistical uncertainty, combined with a longer time extent. Longer time extents are already present in current state-of-the-art lattice simulations, but the required decrease of the error is not yet realistic.


\end{document}